\documentclass[journal,transmag]{IEEEtran}
\UseRawInputEncoding
\usepackage{cite}

\ifCLASSINFOpdf
   \usepackage[pdftex]{graphicx}
   \DeclareGraphicsExtensions{.pdf,.jpeg,.png}
\else
\fi
\usepackage{amsmath}
\usepackage{algorithmic}

\usepackage{array}

\usepackage{stfloats}
\usepackage{url}
\usepackage{doi}
 \usepackage[version=4]{mhchem}

\usepackage{xcolor}
\usepackage{multirow}
\usepackage{hyperref}
\usepackage{caption}
\usepackage{longtable}
\usepackage[acronym,nomain,nonumberlist]{glossaries}

\usepackage{glossary-mcols}
\makeglossaries
\setglossarystyle{listdotted}
\usepackage{tabularx, booktabs} 

\begin{document}
%

\title{Progress towards machine learning methodologies for laser-induced breakdown spectroscopy with an emphasis on soil analysis}


\author{\IEEEauthorblockN{Yingchao Huang\IEEEauthorrefmark{1},
Sivanandan S. Harilal\IEEEauthorrefmark{2}, Abdul Bais\IEEEauthorrefmark{1}, and Amina E. Hussein\IEEEauthorrefmark{3}}
\IEEEauthorblockA{\IEEEauthorrefmark{1} Faculty of Engineering and Applied Science,
University of Regina, SK S4S 0A2, Canada}
\IEEEauthorblockA{\IEEEauthorrefmark{2} Pacific Northwest National Laboratory, Richland, WA 99354, United States}
\IEEEauthorblockA{\IEEEauthorrefmark{3} Department of Electrical and Computer Engineering, University of Alberta, Edmonton AB T6G 2H5, Canada}

\thanks{ Corresponding author:  (email: hari@pnnl.gov, aehussein@ualberta.ca, abdul.bais@uregina.ca, huang47y@uregina.ca).}}

\markboth{XXX,~Vol.~X, No.~X, Month~YYYY}%
{Huang \MakeLowercase{\textit{ \emph{et al.}}}: Applications of laser-induced breakdown spectroscopy for soil analysis: A review}
%



\IEEEtitleabstractindextext{%
\begin{abstract}
 Optical emission spectroscopy of laser-produced plasmas, commonly known as laser-induced breakdown spectroscopy (LIBS), is an emerging analytical tool for rapid soil analysis. However, specific challenges with LIBS exist, such as matrix effects and quantification issues, that require further study in the application of LIBS, particularly for analysis of heterogeneous samples such as soils. Advancements in the applications of Machine Learning (ML) methods can address some of these issues, advancing the potential for LIBS in soil analysis. This article aims to review the progress of LIBS application combined with ML methods, focusing on methodological approaches used in reducing matrix effect, feature selection, quantification analysis, soil classification, and self-absorption. The performance of various adopted ML approaches is discussed, including their shortcomings and advantages, to provide researchers with a clear picture of the current status of ML applications in LIBS for improving its analytical capability. The challenges and prospects of LIBS development in soil analysis are proposed, offering a path toward future research. This review article emphasize ML tools for LIBS soil analysis that are broadly relevant for other LIBS applications. 
\end{abstract}

}

\maketitle

\IEEEdisplaynontitleabstractindextext

%
\IEEEpeerreviewmaketitle

\section{Introduction}
Soil plays a vital role in the Earth’s ecosystem. It provides anchorage for roots and holds water and nutrients. The proper management of soils can only be achieved by the knowledge of soil chemical, physical and biological properties, including nutrient content, organic matter content, pH, texture, bulk density, water retention capacity, and microbial activity \cite{TEJADA200622}. Soils are highly heterogeneous, and thus complex  to characterize.  Resolving its properties is often labor intensive and requires several techniques to extract chemical elements. \par

Traditional analytical techniques, such as Inductively Coupled Plasma-Mass Spectrometry (ICP-MS), ICP-Optical Emission Spectroscopy (ICP-OES), and Flame Atomic Absorption Spectrometry (FAAS), have been widely used for the multi-elemental analysis of soils \cite{NICOLODELLI201970}. However, these conventional techniques are generally complex and laborious, requiring numerous time-consuming steps for sample collection, preparation, and chemical treatment of samples prior to the analysis conducted off-site in a laboratory \cite{PENG2016260}. Advancements in smart agriculture demand low-cost, real-time automated methods for soil sampling directly from the field, and digitized advancement of agricultural data. Therefore, expedited techniques are necessary to provide regular monitoring and rapid evaluations of soil properties, even in real-time, and could ultimately improve soil management and reduce agroecosystem environmental impacts \cite{https://doi.org/10.1111/ejss.12888}. \par

Among the emerging techniques, Laser-Induced Breakdown Spectroscopy (LIBS) has a high potential for soil analysis. LIBS is a type of atomic emission spectrometry based on plasma generation by high-power laser pulses \cite{musazzi2014laser, LIBS-book-singh, Miziolek2006Book}. The advantages of a LIBS instrument over conventional soil sampling techniques are numerous. LIBS employs a single laser shot to obtain spectroscopic data containing multi-elemental measurements on the order of microseconds. It enables simultaneous analysis of many elements with little or no sample preparation compared to the traditional laboratory-based tests. The LIBS technique is also non-invasive, removing only a few nanograms of material without disrupting the larger area. Multiple measurements can be taken to improve statistics and data collection, with data collection rate set by the repetition rate of the laser system, which is typically a matter of seconds. Further, a specific feature that makes LIBS attractive for analyses in the field is the availability of system configurations enabling the construction of portable equipment for in-field measurements \cite{RAKOVSKY2014269}. LIBS is currently used for a wide range of applications, including environmental, space missions, nuclear forensics, isotopic analysis, and steel industry for metal analysis.\cite{musazzi2014laser, LIBS-book-singh, Miziolek2006Book}  LIBS technology has also been implemented on Mars rover Curiosity with ChemCam \cite{https://doi.org/10.1029/2017JE005467} and Perseverance with SuperCam \cite{2021DoSS} for remote analyses of rocks and soils.   \par

With the advancement of the LIBS technique since the 1960s, extensive research efforts have been carried out on the utility of LIBS for soil analysis, and there exist several reviews on this topic with an emphasis on quantitative analysis of soils. For example, Senesi\emph{et al.} reported LIBS and laser spectroscopy applications to the quantitative measurement of carbon content in soil \cite{SENESI20167} and the Humidification Degree (HD) of Soil Organic Matter (SOM) \cite{Senesi2016LaserbasedSM}. Villas-Boas\emph{et al.}\cite{VILLASBOAS2016195} examined the applications of LIBS in determining the soil carbon, nutrients, toxic elements, and classifications. Yu \emph{et al.}\cite{YU2020127} reviewed LIBS applications for soil elemental analysis, such as major, minor nutrient, and heavy metal elements.  \par 

However, while LIBS has made tremendous progress in soil analysis in recent times, there exist several challenges to its use for comprehensive soil characterization, including matrix effects, issues related to quantification, and classification. The most common challenges associated with the application of LIBS for soil analysis are summarised below:
\begin{itemize}
\item \textbf{Trace element detection:} The concentration of trace elements, including micronutrients and toxic elements in soils, is very small, leading to weaker emission line intensities in the LIBS spectrum as compared to strong spectral emission features from the bulk elements and hence cannot be easily detected. For most toxic metals, the LIBS detection limit is between 1 and 20~ppm in a soil matrix \cite{Cremers2013Book}, which cannot meet the requirements of the control of soil pollution \cite {C8RA07799A}, even with the microwave and Laser Induced Fluorescence (LIF) assistance \cite {YiRongxing2018DoTA}. 
\item \textbf{Signal fluctuation:} By taking the average value of LIBS spectral data from several laser shots, the problem of the signal fluctuation in the spectral data can be solved to some extent, but improving the stability of the spectral signal with different morphology is still a problem with quantitative detection \cite{refId0}.
\item\textbf{Soil matrix components difference from different regions:} The characteristic spectrum of the plasma affected by the matrix effect can result in a large difference in LIBS spectra between different soils. Most studies do not consider large sets of samples with varying origins and textures \cite{https://doi.org/10.1111/ejss.12889}. The changing (and unknown) matrix of soil at each site, and variable grain size, are cited as potential issues for the measurement. 
\item \textbf{Weak detection sensitivity:} The detection sensitivity, which restricts the application of LIBS in both qualitative and quantitative analyses, is an important factor and needs to be improved for LIBS detection of minor or trace elements in samples \cite{10.3389/fphy.2020.00068}. For the detection of elements in most solids, the LOD of LIBS is 1 \textendash 100 parts per million \cite{TAKAHASHI201731}, which cannot meet the demands for many trace substances.
\item \textbf{Instrumental challenges:} Challenges associated with LIBS instrumentation include the control of ablation and plasma formation, evolution and interaction with the surrounding environment, the achievement of equilibrium conditions in the plasma, and the elimination of self-absorption effects that may result in a reduction of the signal intensity for elements at high concentrations \cite{SENESI20167}. 
\item \textbf{Comprehensive database for soil:} A large number of analyzed soils in different conditions would be required to build a comprehensive database (allowing for the general model to be built) to compensate for matrix effects. \cite{article11}.
\item \textbf{Feature engineering and selection:} Feature engineering/selection is one of the critical phases of LIBS analysis where key features are related to the element concentration. This process is generally performed either manually or automatically by someone familiar with the LIBS process. However, the sample preparation, element spectra, and data collection methodology varies from model to model, requiring sophisticated understanding of instrumentation and samples. 
 
\end{itemize} 

By overcoming these challenges, LIBS holds the potential to become a viable rapid analysis tool for soil scientists, agricultural producers, agricultural companies, and environmental regulatory agencies. The realization of such a tool, capable of real-time, waste free analysis could greatly benefit these industries by providing regular and readily available information to increase agricultural production and improve farming practices while minimizing environmental impacts \cite{https://doi.org/10.1111/ejss.12889}. 

Over the last decade, machine learning algorithms and data science tools have been used to improve the capabilities in a wide range of engineering, physical, chemical, and biological fields including the capabilities of various analytical tools. In particular, ML methods have provided a vast improvement in the application of LIBS, particularly in soil analysis \cite{YU2020127}. This merger of ML with analytical tools is a novel, rapidly evolving and growing field that contributes to the capabilities of traditional methodologies, where data analysis (peak selection, intensity, etc.)  was performed solely by analysts for making conclusions. Further, the application of ML algorithms to various domain-specific research challenges has in part enabled a paradigm shift from empirical modeling towards data-driven machine learning approaches \cite{article99}. Traditionally, analysis of LIBS spectra is performed by trained analysts considering single or few-parameter correlations, but with the advent of ML tools, more information from the collected spectra can be derived, improving analytical merits, multi-parameter contributions and classification. Additionally, the application of ML to LIBS spectral data allows one to make decisions or conclusions towards optimization of the process, without human-in-the-loop processing during data collection. \par

\begin{figure*}[htbp]
\centerline{\includegraphics[width=\textwidth]{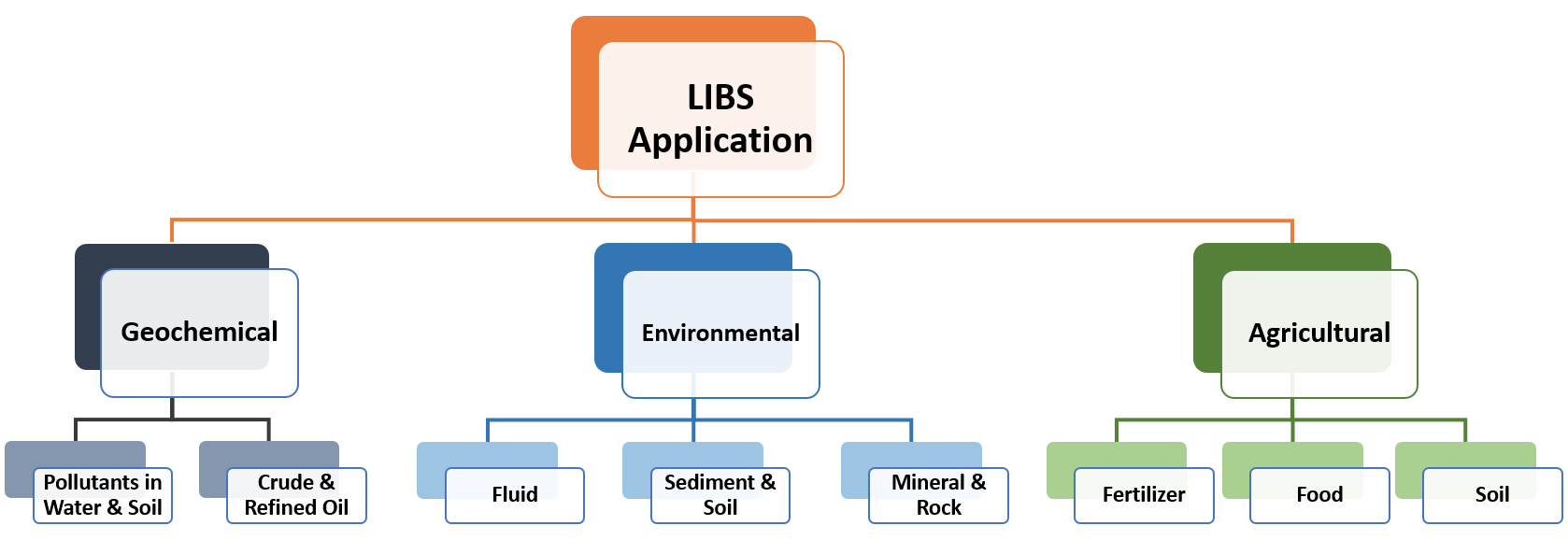}}
\captionsetup{justification=centering,margin=2cm}
\caption{LIBS application areas in which ML methods can be used.}
\label{libsapplication_fig}
\end{figure*}

ML combined with LIBS has been applied in many fields, as shown in Figure~\ref{libsapplication_fig}.  For example, Chen \emph{et al.} \cite{CHEN2020116113} reported that the fusion of LIBS and ML techniques on geochemical and environmental materials, which include seawater, rocks, fossils, coals, and soils, becomes an effective means to improve the accuracy of the classification and quantitative analysis of the information derived from LIBS spectra data sets. A recent review article \cite{LI2021106183} focused on the fusion of LIBS and Artificial Neural Networks (ANN) in different areas, including geology, industry, and biology, highlighting the potential for these techniques to enable enhancements in material identification/classification and component concentration quantification in these fields. The LIBS data processing and analysis are multivariate, nonlinear, nondeterministic, and hence very complicated, which makes the ANN a suitable tool for LIBS investigation as it is capable of dealing with uncertainties, noisy data, and non-linear relationships \cite{1997Neural}. In particular, the performance of ANN techniques, including Back-Propagation Neural Network (BPNN), Radial Basis Function Neural Networks (RBFNN), Self-Organizing Map (SOM), and Convolutional Neural Networks (CNN) are compared \cite{1997Neural}. With more attention to the application of ML in the analysis of LIBS data, Zhang \emph{et al.} \cite{doi:10.1080/05704928.2020.1843175} reviewed the basic principles of LIBS and machine learning algorithms, focusing on Principal Component Analysis (PCA) and Support Vector Machine (SVM), and evaluated their performance on LIBS classification and regression problems. These previous studies highlight the enormous potential of ML approaches in LIBS, and motivate a deeper study of ML techniques specifically applied to soil analysis.\par

The present review article focuses on the potential applications of ML methods for LIBS in soil analysis, including experimental optimization, quantitative measurement, and qualitative identification with additional details of approaches, methodology, and techniques adopted to address current challenges. The review starts with a brief introduction of the LIBS instrumentation and experimental details (section \ref{instrument}), which discusses the optimization challenges of LIBS signal caused by the complex laser\textendash matter interaction and subsequent plasma formation influenced by selected experimental parameters. The matrix effect is regarded as one of the greatest challenges in LIBS in the field of soil analysis, and hence a detailed description of methods to reduce the matrix effect, both using experimental parameters and data analysis, is provided in section \ref{ML}A. Because of the high dimensions in LIBS complex spectral features, which cause inaccurate predictions, feature extraction using  ML methods is an important topic. Section \ref{ML}B details ML methods used in spectral extraction/selection. LIBS has been widely used in soil analysis for elemental measurement and discrimination. Section \ref{ML}C gives an overview of Artificial  Intelligence (AI)/ ML techniques used for quantitative and qualitative analysis. Section \ref{ML}D discusses various ML methodologies used for accurate LIBS classification. A brief discussion of self-absorption in LIBS spectral features is given in section \ref{ML}E. Finally, the potential shortcomings and disadvantages of the ML techniques are discussed in section \ref{prospects}, along with future research directions.

\section{LIBS instrument and experiment}
\label{instrument}

The LIBS technique is based on the collection of visible emission from a plasma generated by an intense laser focused on a target surface. By spectrally resolving the emitted light using a spectrograph, elemental, and in some cases isotopic, information about the sample of interest can be gathered \cite{Cremers2013Book, 2018-APR-Hari}. Even though much of the LIBS-related work for quantitative analysis has appeared in the last two decades, the earliest work can be traced back to 1960’s - following the invention of the laser \cite{Brech1961}. The considerable attention for LIBS in recent times is due to the expanded availability of cost-effective, small footprint, reliable lasers  and spectrographs. LIBS is also superior to other analytical techniques due to its experimental simplicity, multi-elemental, rapid analysis and standoff capability, and minimal sample preparation requirement. \par

\begin{figure} 
\includegraphics[width=0.5\textwidth]{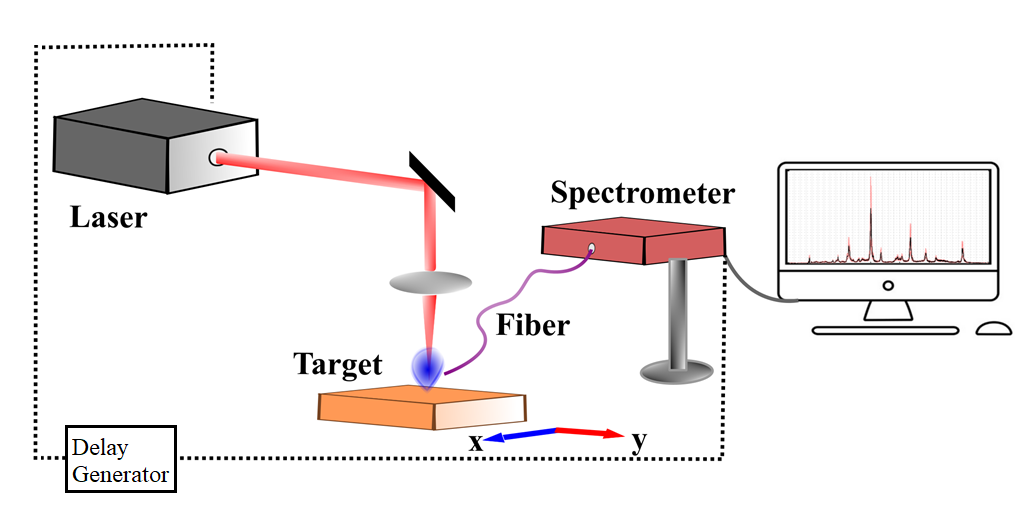}
\caption{\label{libsexperimental} Schematic of LIBS experimental setup.}
\end{figure}

A schematic of the LIBS system is given in Figure~\ref{libsexperimental}. Although the instrumentation looks simple - it contains a high-energy laser, laser focusing, plasma light collection optics, a spectrograph, and a detector for light analysis - the physics involved in the plasma generation and its self-emission is very complex \cite{2018-APR-Hari}. The spectral features from the LIBS plume are contributed by excited atoms, ions, and molecules, and their features vary significantly with time after the plasma onset. The emission properties also depend heavily on several inter-related parameters, \emph{viz.} laser properties (wavelength, pulse duration, energy), focusing conditions (spot size, ablation efficiency), and target properties \cite{2013-JAP-Amina, hahn2012, 2021-JAP-LIz-H-D}. In addition to these considerations of the laser system and plasma generation, the sample environment, such as the nature and pressure of the ambient gas, can influence the plasma properties and hence LIBS spectral features significantly \cite{2020-JAAS-Liz}. Since the parametric space for the LIBS is large, careful optimization is essential for improving the measurement's accuracy and/or sensitivity.\par

The core component is a pulsed laser system with energy sufficient to induce a plasma. The wavelength, repetition rate, and pulse width are the critical considerations for laser selection. The repetition rates are typically in the range 1 to 100 Hz, which can be optimized for maximum power, signal intensity, or the minimization of aerosol production. It has been previously found that LIBS measurements of soil samples exhibited a maximum average signal intensity for lower repetition rates compared to sand samples, due to the production of aerosols at the sample surface \cite{Wisbrun1994DetectorFT}. The most commonly used laser for LIBS is the fundamental radiation (1064~nm) from an Nd:YAG laser with nanosecond (ns) pulse duration. Other candidates are YAG harmonics (532~nm, 355~nm, 266~nm), excimer (193~nm, 248~nm), N$_2$ (337~nm), and CO$_2$ (10.6~$\mu$m) laser systems. IR or NIR ns duration lasers generate hotter plasma conditions at a reduced ablation rate whereas ns-lasers emitting in the UV spectral region provide better laser-target coupling, efficient material removal and vaporization \cite{2013-JAP-Amina}.  In recent times, femtosecond (fs) duration lasers (e.g. Ti:Sapphire) have also been used due to certain advantages over long pulse ns lasers, such as reduced heat-affected zone (HAZ) and lower temperature plasma conditions leading to reduced continuum, generation of an atomic plume and reduced material ablation and smaller craters \cite{LIBS-book-singh, 2018-APR-Hari}. However, ns lasers are still used in most LIBS systems due to their cost-effectiveness, smaller footprint, and easiness of operation (turn-key systems).  \par

\begin{figure}[htbp]
\centerline{\includegraphics[width=\columnwidth]{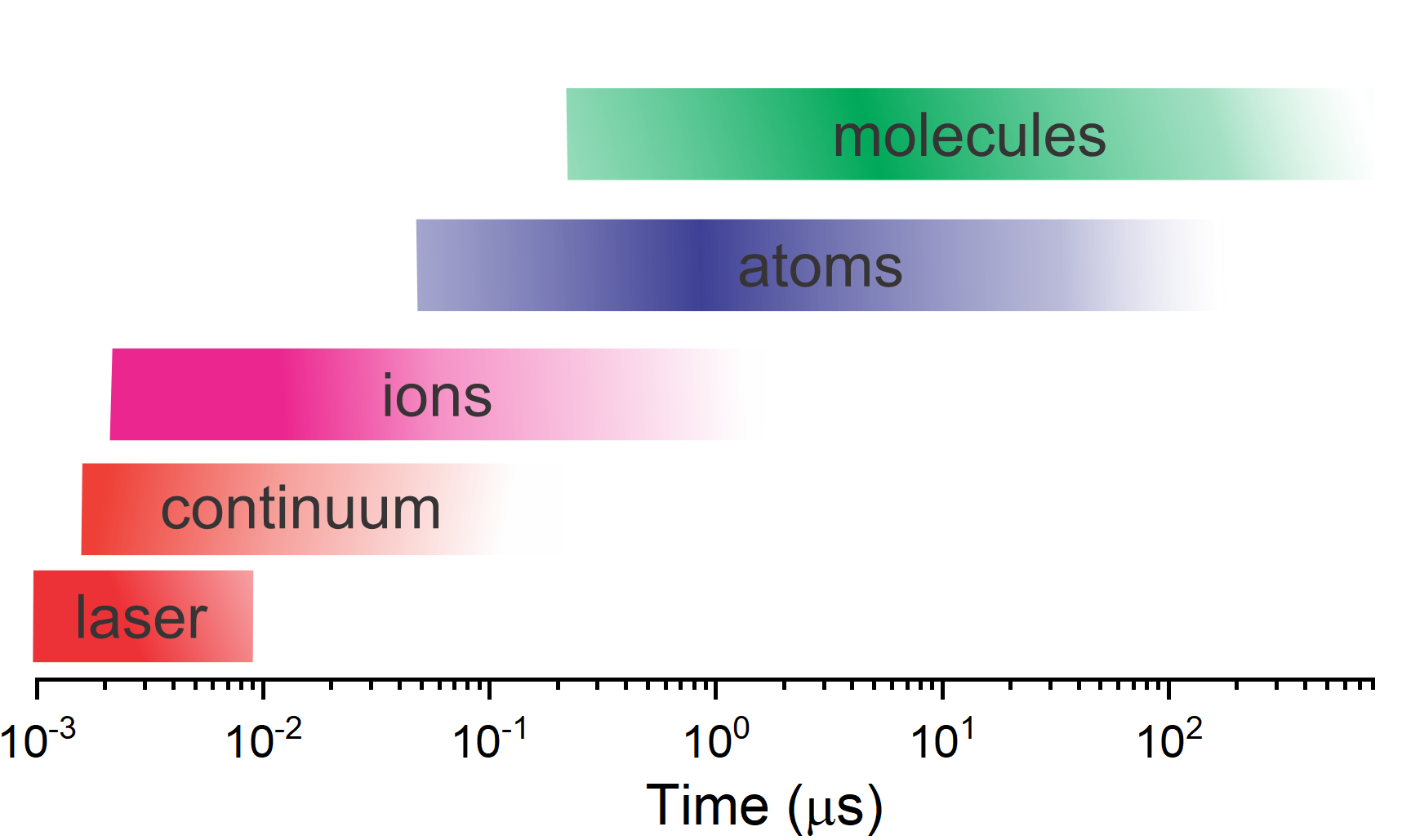}}
\caption{The temporal evolution of emission features observed from a ns LIBS plasma.}
\label{plasmaelapse_fig}
\end{figure}

A combination of a spectrometer and a detector is used for the analysis of light emitted from the plasma, which provides the intensity of line radiation against the wavelength. The properties of the spectrometer determine the throughput, spectral resolution, and bandwidth, while the detector governs the speed of the signal acquisition and the time resolution. Charge-coupled device (CCD) cameras are regularly used for portable spectrographs that provide time-integrated emission analysis, while intensified CCDs (ICCD) provide time resolution and greater sensitivity at lower light levels.  Low-resolution ($\lambda/\Delta \lambda \leq$ 1000) spectrometers are adequate for measuring spectral features of low-Z elements, while medium to high-resolution ($\lambda/\Delta \lambda \geq$  5000) spectrographs are needed for analyzing the complex multi-elemental or high-Z spectral features \cite{2018-APR-Hari}. The excitation characteristics of the plasma (temperature and electron density) are a strong function of the elapsed time after the onset of plasma formation. Hence the emission intensity of ionic, atomic, and molecular transitions vary significantly with time after the plasma onset. Approximate temporal history of continuum, ion, atom, and molecular emission features from a ns-LIBS plume is given in Figure~\ref{plasmaelapse_fig} \cite{doi:https://doi.org/10.1002/9780470027318.a5110t.pub3, 2018-APR-Hari}. Early stages of plasma evolution are marked by strong continuum emission, and it may obscure line emission. The ratio of neutral and ionized species lines can also change as a function of time due to the temporal evolution of electron density and temperature in the plasma. The molecular emission appears at later times of plasma evolution when the plasma temperature is lower \cite{2021-SCAB-LIZ-Review-UO}. Time-gated detection is helpful for discerning continuum and molecular emission features from atomic emission \cite{2018-APR-Hari}. The LIBS spectral features are also useful for measuring physical conditions of the plume \cite{Hari2022review}.   \par

\begin{figure*}[htbp]
\centerline{\includegraphics[width=\textwidth]{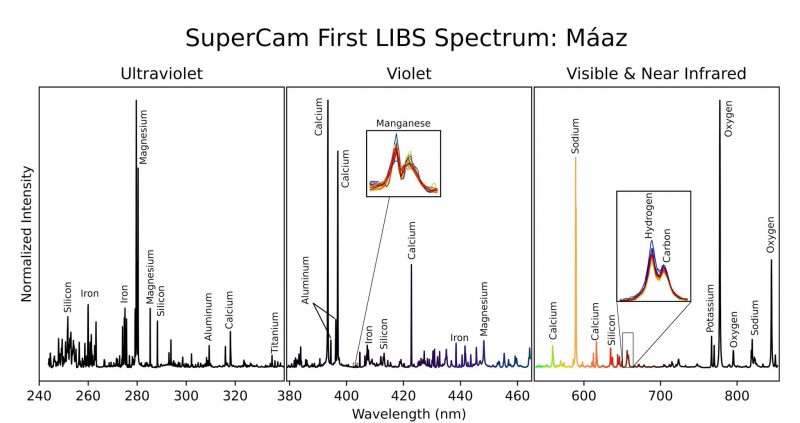}}
\caption{SuperCam first LIBS spectrum on Perseverance showing the composition of the Mars rock target. The rock is basaltic and basaltic and is either igneous (volcanic) or is composed of fine grains of igneous material cemented together by water \cite{earthsky}.}
\label{libsspectra_fig}
\end{figure*}

 Each element in the periodic table is associated with unique LIBS spectral peaks, which can be used to identify the composition of the target or quantify the concentration of trace and major elements in the sample through spectroscopic analysis. The atomic spectral information of various elements is documented in several databases, including the National Institute of Standards and Technology (NIST)\cite{NIST_database}, and Kurucz database \cite{Kurucz_database}. Due to sample heterogeneity, LIBS spectra of soil samples are complex, with thousands of emission lines in which the emission line of one element may interfere with another \cite{D0JA00026D}. Figure~\ref{libsspectra_fig} illustrates LIBS Spectrum from SuperCam on Perseverance showing the composition of the Mars rock target with main emission lines standing for different elements which are identified accordingly \cite{earthsky}. \par  

LIBS is an excellent tool for qualitative and semi-quantitative elemental analysis based on the analysis of spectral emission lines. The detection limits of LIBS vary significantly among the elements in the periodic table and are often at the part-per-million level for many elements \cite{Cremers2013Book}. However, there exist several challenges to accurate quantitative analysis using LIBS. The challenges are relatively lower sensitivity compared to other lab-based tools (e.g., mass spectrometry) \cite{MeissnerK2004Aotm}, matrix effects, congested spectra for complex and high-Z targets, quantification, and data reproducibility. For quantitative analysis, matrix-matched calibration standards are regularly used \cite{TAKAHASHI201731}. \par

Several methods have been proposed to improve LIBS's detection limits, such as double-pulse LIBS \cite{dual-pulse-LIBS}, microwave heated LIBS \cite{microwave-LIBS}, spark excitation \cite{Spark-LIBS}, nanoparticle enhanced LIBS \cite{DEGIACOMO201419}, and LIF of LIBS \cite{2018-OL-Hari-standoffLIF}. Additionally, variations in the signal intensity due to shot-to-shot fluctuations is a major concern for soil analysis. Potential sources for LIBS emission signal shot-to-shot fluctuations include changes in the  laser-sample or laser-plasma interaction and subsequent changes in plume morphology, shot noise due to the number of photons arriving on the detector, etc.\cite{2021Fu-LIBS-fluctuation, SNR-LIBS-Cristoforetti} \par

Currently, a large parametric space is explored in LIBS experiments to mitigate challenges and optimize  measurements. However, the selection of the experimental parameters and hence the optimization of the LIBS system can also be performed with the help of ML methods. For example, Belloua \emph{et al.}\cite{BellouElli2020Lbsa} investigated the impact of experimental parameters such as laser energy and the temporal gating conditions (i.e., delay time and integration time) of the CCD detector of the spectrometer on the plasma characteristics and subsequently on the classification using PCA and Partial Least Squares (PLS). Their work demonstrated that LIBS could be experimentally optimized for the classification of olive oil using by machine learning algorithms An Artificial Neural Network (ANN) was investigated in another recent work to find the optimal experimental laser pulse and gate delay parameters in predicting the signal-to-noise ratio (SNR) of selected spectral lines. The input data for the ANN model comprised not only information about experimental parameters, but also the selected spectral line and matrix parameters. They found that the model is not matrix-independent or universal for all spectral lines, but the number of optimization measurements is reduced dramatically \cite{D1JA00389E}. \par

\section{Methodological and ML approaches for LIBS soil analysis}
\label{ML}

\begin{figure*}[htbp]
\centerline{\includegraphics[width=\textwidth]{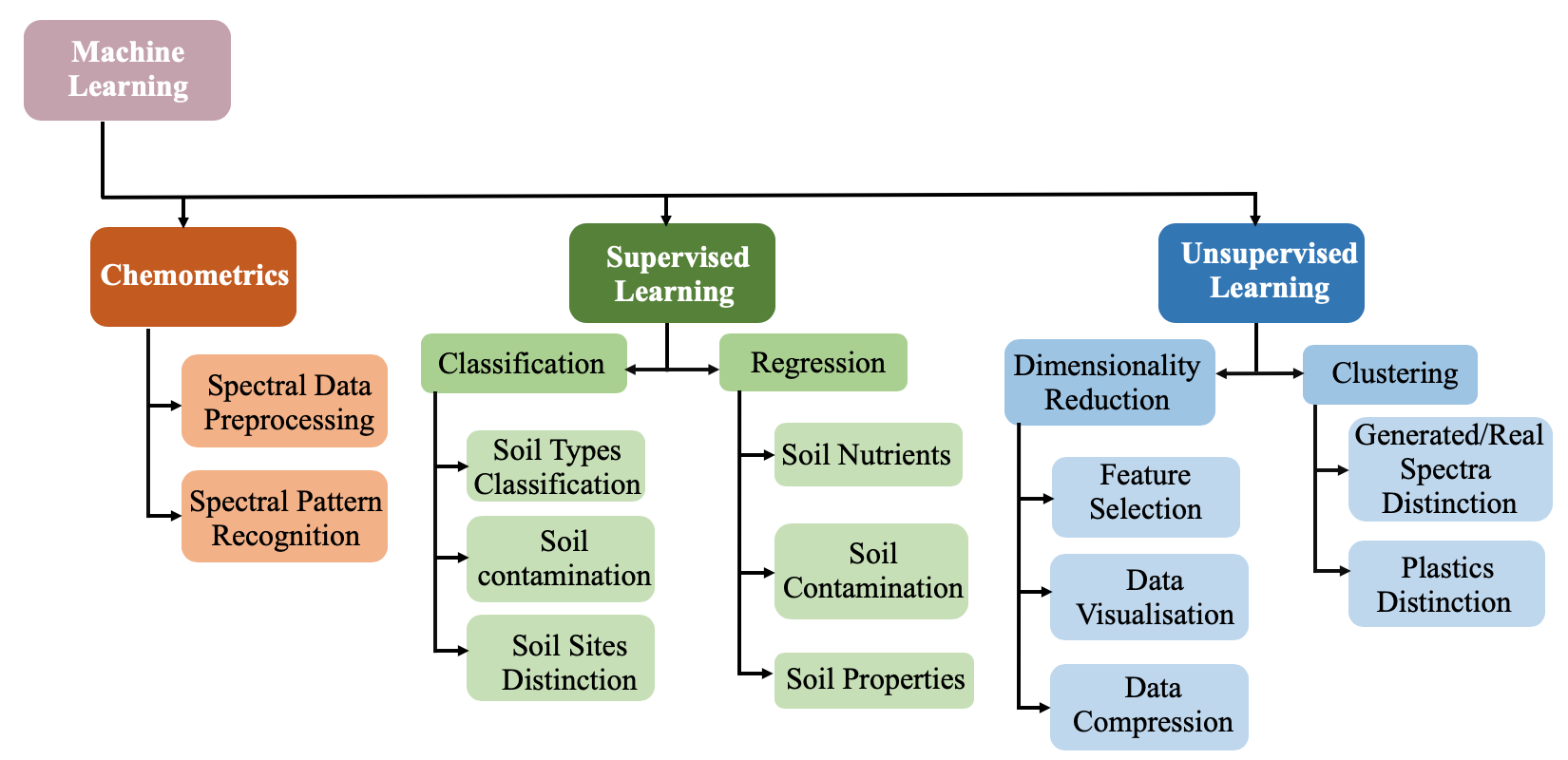}}
\caption{Machine Learning approaches and application for LIBS data analysis.}
\label{ml_fig}
\end{figure*}

Analysis of a LIBS spectrum typically consists of pre-processing to reduce matrix effects, feature extraction/selection to better represent the elements in samples, and classification or quantitative analysis of a sample of interest. Machine learning (ML) is a branch of AI, which consists of many other sub-fields such as computer vision, expert systems, and robotics etc., mainly focusing on the use of data and algorithms to imitate the way that humans learn, gradually improving its accuracy. Through statistical methods, algorithms can be trained to make classifications or predictions, uncovering critical insights within data mining projects. ML techniques have been used in many applications with LIBS, as illustrated in Figure~\ref{ml_fig}. \par

Although ML is an emerging tool for LIBS spectral data analysis, it should be noted that the chemometric methods have been around for some time for spectral data processing and analysis. Chemometrics is a set of mathematical tools and statistical methods, which can be seen as a subset of the ML domain, mainly applied to linear problems. In contrast, ML can be summarized as a set of advanced mathematical and statistical methods for complex issues, in particular for non-linear problems. Chemometrics has a long history which can be traced back to 1960s with some earlier applications in nuclear magnetic resonance (NMR) spectroscopy chemical pattern recognition \cite{chemometrics_history}. It enables multivariate data analysis and offers advantages in data processing, signal analysis, and pattern recognition. Chemometrics methods have been applied to LIBS for spectral preprocessing as well as for qualitative and quantitative analyses \cite{https://doi.org/10.1002/cem.2983}. \par

With the development of ML techniques, both unsupervised and supervised learning have been applied in LIBS analysis. Unsupervised learning is used in dimensionality reduction and clustering. Supervised learning can be used for classification with predefined labels assigned to spectra. Regression under supervised learning is also popular in LIBS analysis. Various ML methods are discussed in the following sections, along with the details about the conventional methods.

\subsection{Matrix effect reduction techniques}
The sample matrix is a major factor affecting the quantitative accuracy of any analytical tool, including LIBS \cite{matrix}. \textit{Matrix effects} are due to changes in emission line intensities of some elements caused by variation in the physical properties and/or the chemical composition of the sample material. These effects primarily influence the production of atomized material by ablation and the excitation properties of the LIBS plume \cite{Cremers2013Book}. \par

There are two kinds of matrix effects: physical and chemical. The so-called \textit{physical} matrix effects are related to the physical properties of the sample surface (e.g., structure, grain size, texture, reflectivity, moisture, morphology, and hardness). These effects may alter the amount of ablated mass, which may cause a variation of the line emission intensity even if the concentration is the same in the various matrices. The \textit{chemical} matrix effects occur when the presence of one element alters the emission behavior of another element because of changes in the amount of ablated material of the various elements due to differences in heat capacity, vaporization temperatures and ionization in the plasma \cite{SENESI20167}. For example, an element present in an equal concentration in two different host materials may exhibit different LIBS emission intensities due to the combined effect of other components. These effects make it very challenging to find matrix-matched standards to perform quantitative analysis of natural samples. A summary of different matrix-effect reduction methods is given in Table~\ref{MatrixEffectTable} and discussed in detail below. \par

\begin{table}[htbp]
\caption{Matrix effect reduction methods used in LIBS spectral analysis}
\begin{center}
\begin{tabular}{>{\centering\arraybackslash}m{1.5cm}|>{\centering\arraybackslash}m{3cm}|>{\centering\arraybackslash}m{1.5cm}}
\hline
\textbf{Methodology}&\textbf{Method}&\textbf{References} \\

\hline
&Measurements made at low ambient pressure&\cite{doi:10.1177/0003702816664858}\\
\cline{2-3}
&Combination of LIBS with LIF&\cite{doi:10.1021/acs.analchem.6b03969},\cite{Gao:18}\\
\cline{2-3}
By Experiment &Work parameters (e.g. laser energy and delay time) optimization&\cite{refId0}\\
\cline{2-3}
&Repeated measurements&\cite{DIAZ:12},\cite{RUHLMANN2018115}\\
\cline{2-3}
&UV laser excitation&\cite{ZAYTSEV201865}\\
\cline{2-3}
&CF-LIBS&\cite{article6},\cite{Kwak2012DeterminationOL},
\cite{CORSI2006748},\cite{2010ApPhB..98..231P},
\cite{B820493D},\cite{C6RA13038K}\\
\hline

&Wavelet Transformation&\cite{article5},\cite{soilsystems3040066},\cite{WANG2018300}, \cite{https://doi.org/10.1002/cem.2422}\\
\cline{2-3}
&Normalization of peak intensity with background signal&\cite{Kwak2012DeterminationOL},\cite{DIAZ:12}\\
\cline{2-3}
&Multiple-shot averaging&\cite{article5},\cite{ZAYTSEV201865}\\
\cline{2-3}
&Inclusion of matrix information (e.g. type of soil and sample preparation method) with spectra intensities in the inputs&\cite{sun:hal-02276189}\\
\cline{2-3}
&Discarding method&\cite{C3JA50233C}\\
\cline{2-3}
Data&Internal Standardization&\cite{KIM2013754}\\
\cline{2-3}
Treatment&Spectral smoothing&\cite{VILLASBOAS2016195},\cite{doi:10.1021/acs.energyfuels.7b01718}\\
\cline{2-3}
&Background subtraction, Gaussian denoising, continuum removal and wavelength adjustments&\cite{castorena2021106125}\\
\cline{2-3}
&Spectra elimination with Euclidean distance to the center of the PC&\cite{s20020418}\\
\cline{2-3}
&Offset correction&\cite{Nicolodelli:14},\cite{VILLASBOAS2016195}\\
\cline{2-3}
&Alignment and baseline correction&\cite{RUHLMANN2018115},\cite{ZAYTSEV201865},\cite{VILLASBOAS2016195},
\cite{WANG2018300},\cite{LILAND2015135},\cite{doi:10.1366/000370210792434350},
\cite{Liland2011OptimalBC},\cite{soilsystems3040066}\\
\hline
\end{tabular}
\label{MatrixEffectTable}
\end{center}
\end{table}

One of the methods to reduce the matrix effects is repeated LIBS measurements and data averaging. Díaz \emph{et al.} \cite{DIAZ:12} evaluated the repeatability of the sampling method, collecting three consecutive 100-shot spectra for each sample and using ensemble-averaged spectra as the average of 100 successive single shots to decrease dispersion. Similarly, Rühlmann \emph{et al.} \cite{RUHLMANN2018115} repeated measurements at different positions of each sample and averaged the measurements to reduce the effect of sample heterogeneity. He \emph{et al.} \cite{article5} and Zaytsev \emph{et al.} \cite{ZAYTSEV201865} considered the accumulated average of multiple spectra to eliminate the apparent matrix effect. \par

Additionally, the background in LIBS spectra may vary from one spectral region to another because emission spectra are strongly affected by the acquisition system, which depends on optical and electronic systems and may impose offset and random noise on the spectra acquired. To reduce random noise, the spectra may be smoothed using a Savitzky\textendash Golay filter whose parameters were selected according to a genetic optimization algorithm. For example, Nicolodelli \emph{et al.} \cite{Nicolodelli:14} corrected the offsets in their measurements by subtracting the average of the random noise region close to the analyzed carbon element emission line. The Wavelet Transformation (WT) is also a valuable methodology in modeling data characterized by sharp peaks, spikes, and other local features using a set of wavelet basis functions \cite{nason2010wavelet}. As an efficient denoising method, the WT approach is used to reduce the noise generated by the instrument and the sampling process \cite{s17081894}. In addition, WT is used to minimize irregularities in the LIBS spectra before elemental analysis in many studies \cite{article5}\cite{soilsystems3040066}\cite{https://doi.org/10.1002/cem.2422}.  \par

\begin{figure*}[htbp]
\centerline{\includegraphics[width=\textwidth]{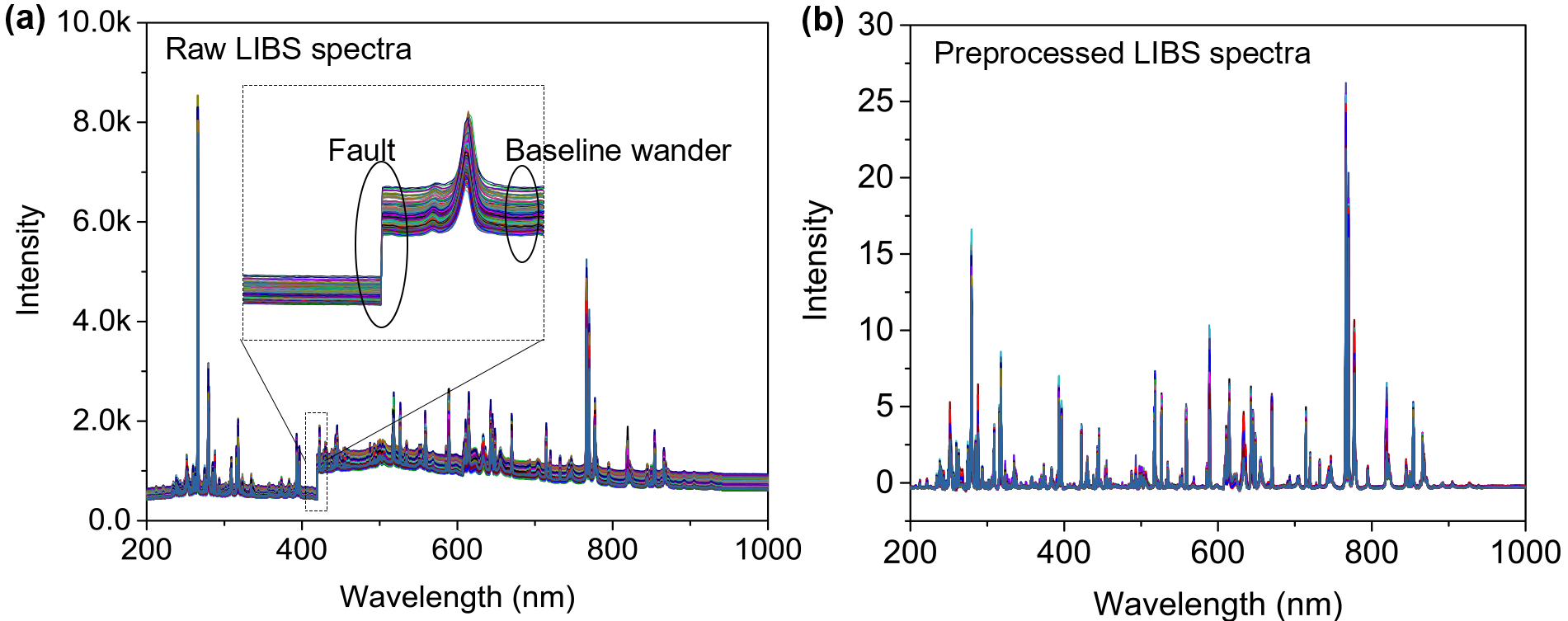}}
\caption{(a) Raw spectra (b) preprocessed spectra with baseline correction \cite{XU2019113905}.}
\label{matrixeffect_fig}
\end{figure*}

It is often more effective to remove an estimated baseline from the data when measuring many peaks by including a baseline function. The baseline is a recurring problem because of the strong residual background emission caused by congested spectra, reduced spectral resolution, continuum, molecular emission, and thermal emission from nanoparticles \cite{2021-SCAB-LIZ-Review-UO}. The baseline can be corrected for a broad spectral range with an algorithm that fits a smooth curve to the baseline points that are in regions free of well-defined peaks  \cite{doi:10.1366/000370210792434350,Liland2011OptimalBC,LILAND2015135}. Villas-Boas \emph{et al.} \cite{VILLASBOAS2016195} corrected the baseline in their experimental measurements of chemical elements within soils through estimation of the baseline by iteratively suppressing the spectrum with a moving window. Zaytsev \emph{et al.} \cite{ZAYTSEV201865} also removed the baseline in determining lead in soils using an iterative algorithm described by Torres \emph{et al.} \cite{TorresE.L1998Ssft}. Wang \emph{et al.} \cite{WANG2018300} detected and removed the baseline by iteratively performing a polynomial fitting in the spectrum. Morphological Weighted Penalized Least Squares (MPLS) is another option to correct the baseline \cite{soilsystems3040066}. Li \emph{et al.} \cite{C3AN00743J} provided a  detailed theory of the MPLS algorithm. An example of pre-processed data using the baseline correction approach is given in Figure~\ref{matrixeffect_fig}. The baseline wander and fault in the raw spectra could be observed, and these features are obviously eliminated with baseline correction. \par

Normalization is the most widely used processing method to reduce matrix effects. Here an emission line is typically normalized to the intensity of a specific element (i.e., the standard normalization method) for minimizing the shot-to-shot variation and matrix effects seen in soil samples. Kwak \emph{et al.} \cite{Kwak2012DeterminationOL} normalized the peak area of lead by major element emission lines, which gave the highest linearity with concentrations. Díaz \emph{et al.} \cite{DIAZ:12} applied normalization using total background intensity that is generally restricted to online control processes where the sample’s matrix is almost invariable. The peak-to-base normalization technique ratio was used to decrease the effect of plasma variations from shot-to-shot and sampling artifacts present when analyzing heterogeneous samples. Kim \emph{et al.} \cite{C3JA50233C} applied a discarding method as the first step in data analysis by deleting the highest and lowest 10\% of the total data from 100 peak areas in each soil. Then, the study peak areas of zinc were normalized using four different methods, including SNR, Signal to Background Ratio (SBR), Carbon Normalization (SCR), and Full Spectrum Normalization (SFR), which minimized factors affecting the LIBS result.\par

Experimentally, selection of the appropriate environment (e.g. nature and pressure of the ambient conditions) is an effective way to reduce the matrix effects. A low-pressure cover gas provides a reduced matrix effect and is found to provide higher accuracy in obtaining calibration curves \cite{doi:10.1177/0003702816664858}. This is because chemical matrix effects come from a strong interaction between atoms in proximity, such as high cover gas pressure conditions yielding dense plasma due to plasma confinement. Laser Induced Fluorescence (LIF) of the LIBS plume is also an effective method for reducing the matrix effects, where preferential excitation can boost the analyte signal intensity. A LIF/LIBS combination has previously been used for detecting heavy trace elements in soils \cite{doi:10.1021/acs.analchem.6b03969}\cite{Gao:18}. Several authors \cite{ZAYTSEV201865,UV-matrix} have also tried to reduce the matrix effects using ultraviolet laser radiation for plasma generation \cite{ZAYTSEV201865}.\par

Calibration-Free LIBS (CF-LIBS), as the name implies, does not need the use of matrix-matched calibration curves as standards, instead uses the measurement of line intensities and plasma properties (\emph{viz.} excitation temperature and electron density) under the assumption of a Boltzmann population of excited levels.  Ciucci \emph{et al.} \cite{doi:10.1366/0003702991947612} used the CF-LIBS approach, which accounts for both the physical and chemical matrix effects through analysis of the spectrum without the real need for calibration curves. It corrects matrix effects theoretically based on the Boltzmann distribution law and Saha equation \cite{TOGNONI20101, PENG2016260, Kwak2012DeterminationOL, CORSI2006748,  2010ApPhB..98..231P}. Being more theoretical and free from errors associated with the matrix effect, the CF-LIBS is considered a suitable approach for carrying out the quantitative analysis of multi-elemental materials, such as soils \cite {app10196848}. For example, Kwak \emph{et al.} \cite{Kwak2012DeterminationOL} used CF-LIBS to determine Pb concentration in soils. Wang \emph{et al.} \cite{C6RA13038K} combined CF-LIBS with a Binary Search Algorithm (BSA) to determine the acidity (CaO/SiO2 mass ratios) of iron ore. It is noteworthy that BSA is a classical search method and was utilized to search for the optimal plasma temperature in CF-LIBS. In another recent study, the CF- LIBS method was used to conduct a quantitative analysis of the contamination factors of metals in mining sediments \cite{article6}. \par

Internal standardization is another way to minimize the matrix effect. An internal standard in analytical chemistry is a chemical substance added in a constant amount to samples in chemical analysis. It is a compound that must be shown similar behavior to the analyte, which is done to correct analyte losses during sample preparation \cite{article0}. In this approach, calibration curves are developed for an element in the material of interest, and then LIBS measurements for the unknown samples are acquired under identical experimental conditions. Kim \emph{et al.} \cite{KIM2013754} applied the internal standardization method to determine whether the matrix effect introduced by water content in samples could be reduced, suggesting that if the water content is less than 10\%, the impact of soil water content on the LIBS response can be minimized.\par 

The influence of different spectral preprocessing techniques such as smoothing, Standard Normal Variate transformation (SNV), Multiplicative Scatter Correction (MSC), Mean Centring (MC), and derivation by convolution (Savitzky \textendash Golay) are compared on the quantitative model for calorific value analysis with coal samples by LIBS \cite{doi:10.1021/acs.energyfuels.7b01718}. These results show that spectral smoothing effectively reduces the differences of matrix effects among different samples. Different numbers of points were experimented in smoothing and found that the quantitative model obtained the best comprehensive performance with the application of the 11 points smoothing combined with the second-order derivation. The inter-element influences could be further eliminated with the derivative convolution (Savitzky\textendash Golay). \par

The concept of the generalized spectrum was developed by Sun \emph{et al.} \cite{sun:hal-02276189}, where information about the soil matrix (type of soil, sample preparation method etc.) was explicitly included in the input vector of the model as an additional dimension to reduce the matrix effect. The multivariate calibration model was designed to consider the properties of different soils, which efficiently reduced data dispersion due to experimental fluctuations. Spectral variance can also be reduced using the Euclidean distance of a data point to the center of the Principal Component (PC) space to remove a predetermined percentage of spectra with the largest Euclidean distances before averaging the remaining spectra of one sample \cite{s20020418}. Castorena \emph{et al.} \cite{castorena2021106125} conducted spectra preprocessing by background subtraction, Gaussian denoising, continuum removal, and wavelength adjustments to reduce the matrix effect. \par

\subsection{Feature extraction for LIBS spectra}
Feature engineering uses domain knowledge to extract features such as characteristics, properties and attributes from raw data. Overfitting becomes an issue when the LIBS spectral range spans a large wavelength region and contains crowded spectral features, which results in inaccurate predictions by using the whole spectrum for quantitative analysis as some spectral regions may introduce sources of error \cite{2015Less}. Therefore, feature engineering is needed to select the relevant information from the raw spectra to better represent elements of interest. Features are fed into predictive models, and features extracted with different methods influence the test results.\par

There are findings that matrix effects and interference factors could be mitigated by a small fraction of these emission lines using feature selection methods \cite{Nicolini1504146}. It is also showed that the feature selection is effective in avoiding the influence of spectral interference and reducing the influence of self-absorption \cite{Sun_2021}. Therefore, feature engineering is useful to alleviate matrix effects. Table~\ref{featureTable} summarizes the feature selection methods used for LIBS spectral analysis. \par

\begin{table}[htbp]
\caption{Feature extraction methods used for LIBS spectral analysis}
\begin{center}
\begin{tabular}{>{\centering\arraybackslash}m{1.5cm}|>{\centering\arraybackslash}m{3cm}|>{\centering\arraybackslash}m{1.5cm}}
\hline
\textbf{Category}&\textbf{Method}&\textbf{References} \\

\hline
&PCA&\cite{vrabel2020105872},\cite{lu2018},\cite{doi:10.1177/0003702819826283},\cite{ilhardt2019119},
\cite{article4},\cite{article9},\cite{KIM2013754},
\cite{Meng:17},\cite{RUHLMANN2018115},\cite{ZAYTSEV201865},
\cite{s20020418},\cite{CLEGG200979}
\\
\cline{2-3}
By Main Variance&PLS&\cite{soilsystems3040066},\cite{Ferreira14},\cite{VILLASBOAS2016195}\\
\cline{2-3}
&LDA&\cite{vrabel2020105872}\\
\cline{2-3}
&iPLS \& mIPW-PLS&\cite{C7JA00114B}\\
\hline
 
&Spectra segment&\cite{Gu2016}\\
\cline{2-3}
By Spectra Peaks&Features near strongest emission line&\cite{WANG2018300}\\
\cline{2-3}
&By main elements' intensities&\cite{C9AY00890J},\cite{https://doi.org/10.1002/cem.2422}\\
\hline

Deep&CNN&\cite{castorena2021106125},\cite{lu2018},\cite{li2020105850}\\
\cline{2-3}
Learning&MLP&\cite{FERREIRA2011435}\\
\hline

By Evaluation&BestKSelct&\cite{sun:hal-02276189},\cite{ZHANG2020105802},\cite{D0JA00431F}\\
\cline{2-3}
Metrics&FFS&\cite{DANDREA201452}\\
\cline{2-3}
&VI \& VIP&\cite{vrabel2020105872},\cite{LIANG2020104179}\\
\hline
Hybrid&V-WSP \& PSO&\cite{YAN201935}\\
\cline{2-3}
Selection&WT \& MIV&\cite{article2}\\
\hline
\multirow{4}{*}{Others} &Moving window&\cite{vrabel2020105872}\\
\cline{2-3}
&Handcraft feature selection&\cite{vrabel2020105872}\\
\cline{2-3}
&LASSO&\cite{WANG2018300},\cite{doi:10.1366/12-06983},\cite{osti_1582110}\\
\cline{2-3}
&GA&\cite{Li:17}\\
\hline
\end{tabular}
\label{featureTable}
\end{center}
\end{table}

It is also common to extract features from raw data by variance. PCA, PLS, Regression, and Linear Discriminant Analysis (LDA) are the techniques used to catch the main variance of data with high dimensions. PCA is a traditional multivariate statistical method commonly used to reduce the number of predictive variables and solve the multi-collinearity problem \cite{doi:10.1198/016214505000000628}. As a dimension reduction methodology, PCA is applied without considering the correlation between the dependent variable and the independent variables, while PLS is applied based on the correlation \cite{article3}. PCA and PLS have been the dominant methods in the past decade for feature extraction from LIBS spectra. There are reasons for the popularity of PCA and PLS in feature engineering, such as correlated feature removal and visualization improvement. However, these techniques have limitations that make them less suited for LIBS analysis. For example, independent variables become less interpretable with principal components compared to original features. LDA is another statistical method to find a linear combination of features, and it is closely related to PCA. However, instead of an unsupervised learning method like PCA, LDA is a supervised learning algorithm used as a dimensionality reduction technique. In the Euro-Mediterranean Symposium on LIBS (EMSLIBS) contest for soil classification, Schreckenberg \emph{et al.} \cite{vrabel2020105872} used LDA for spectra classification and achieved the second top accuracy for the test set. \par

A modified PLS approach has also been proposed to select variables by combining interval PLS (iPLS) and modified Iterative Predictor Weighting-PLS (mIPW-PLS) for quantitative analysis \cite{C7JA00114B}. In the iPLS algorithm, the spectra were divided into several intervals based on wavelength, and PLS models were generated for each interval. Selecting a few intervals for modeling was found to provide better prediction results than using a single model for the entire spectra. The mIPW-PLS method is based on the cyclic repetition of PLS regression (PLSR). The coupling of iPLS with mIPW-PLS has demonstrated high efficiency in selecting input spectra \cite{C7JA00114B}. On its own, iPLS provides an overview of interesting spectral areas which could be selected and the graphic output gives an overview of the spectral data \cite{doi:10.1366/0003702001949500}. Combined with mIPW-PLS, only variables with maximal relevance are extracted, which further reduces the input dimensions. \par

Since LIBS spectra from soil samples contain intensity peaks from various elements, some researchers select the peak spectra as input features based on known elements in samples, which makes the feature interpretation easier, as compared to PCA which transforms and obscures input variables. In an early study, known wavelengths corresponding to given elements in the spectrum were selected and used in the analysis to aid dimension reduction \cite{https://doi.org/10.1002/cem.2422}. Guo \emph{et al.} \cite{C9AY00890J} selected 14 spectral bands based on known elements in soil samples, including the characteristic lines and the surrounding noise, and then used them as input features. In another work, Gu \emph{et al.} \cite{Gu2016} selected the specific 424 to 428~nm spectral region for chromium (Cr) analysis in soils where only a few atomic emission lines were free of overlap from the heavy metal element Cr, saving the running time due to a reduction in the number of input variables. In a later study, Wang \emph{et al.} \cite{WANG2018300} chose the strongest emission lines of the corresponding elements as variable inputs to aid dimension reductions. However, interference and overlapping between elements make it hard to exclude them when choosing the feature spectra of known elements.\par

Deep learning techniques learn to extract features from data input. Deep Learning is part of a broader family of machine learning methods based on ANN that progressively use multiple layers to extract higher-level features from the raw input. For example, in image processing, lower layers may identify basic features such as edges and textures, while higher layers may identify advanced features that humans can understand, such as digits, letters, or faces. With deep learning techniques, features do not need to be extracted from the spectra manually. Instead, the network learns to select the features in the training process. For example, in an early work, Ferreira \emph{et al.} \cite{FERREIRA2011435} used Multilayer Perception (MLP), an artificial network, to learn the element quantities. In recent works, a convolution neural network (CNN) was proposed and applied with raw LIBS spectra \cite{lu2018, li2020105850, castorena2021106125}. The details of MLP and CNN application are introduced in \ref{ML}C about soil elemental analysis and in \ref{ML}D about classification. \par

\begin{figure}[htbp]
\centerline{\includegraphics[width=\columnwidth]{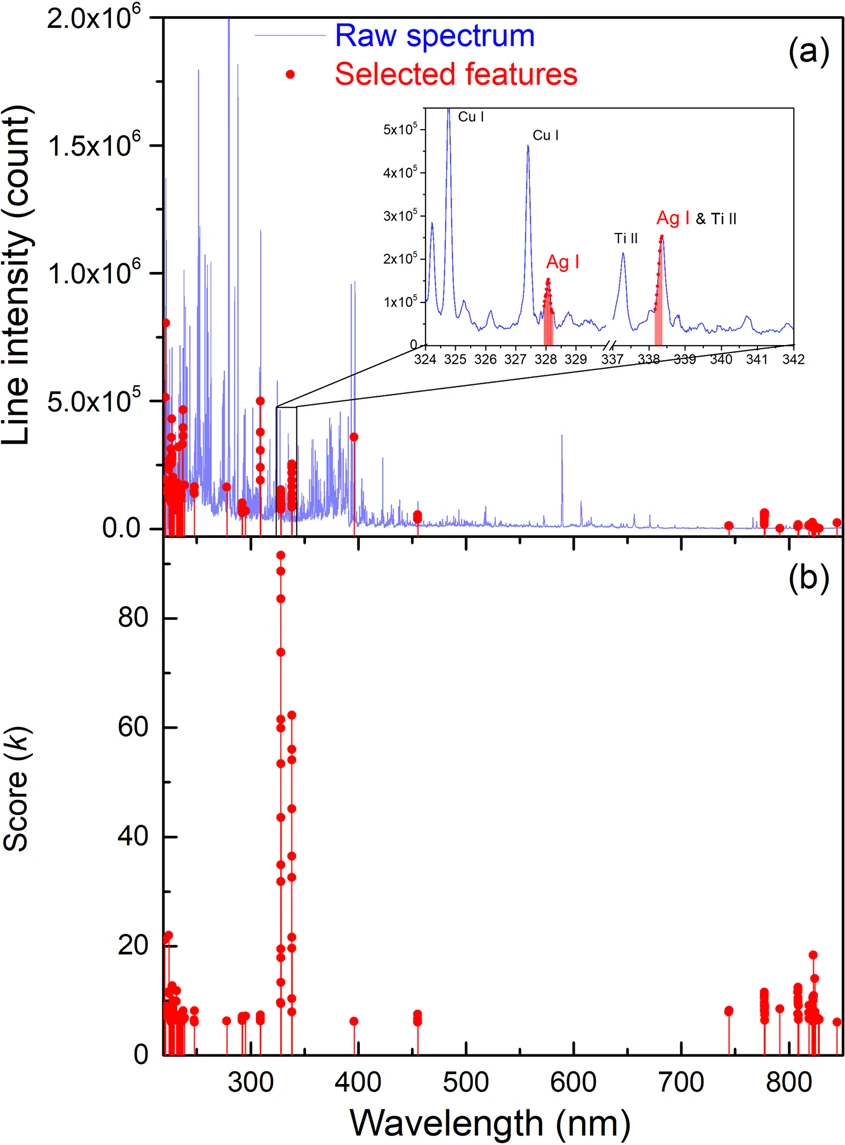}}
\caption{(a) Spectrum of the selected first $k$ features (in red) of the input variables with the highest scores, the raw spectrum (in light blue) is also shown for comparison (b) The SelectKBest scores of selected spectrum (in red) \cite{sun:hal-02276189}.}
\label{featureselection_fig}
\end{figure}

Some evaluation metrics also provide a means of feature extraction. For example, the SelectKBest method selects the features according to the $k$ highest score and it uses analysis of variance (ANOVA) F-values to score all input features. Features selected with SelectKBest are marked with red in Figure~\ref{featureselection_fig} \cite{sun:hal-02276189}. SelectKBest only retains relevant features and drops unwanted features by retaining the first $k$ features of the input variables with the highest scores. The principle is to calculate the covariance between the intensity of a given channel in a spectrum and the element concentration of the corresponding sample, then, only the pixel intensities with high enough correlation with the series of analyte concentrations are selected and kept for further processing \cite{sun:hal-02276189}\cite{ZHANG2020105802}\cite{D0JA00431F}. 

The impact of a given variable on the desired output/analysis can also be used as an evaluation metric in SelectKBest. Forward Feature Selection (FFS) is an evaluation technique that selects the features that have the most significant effect on the output, such as the highest accuracy. FFS initially scans all feature subsets comprising only one feature and selects the best one. Then, FFS searches the best subset including two features, three features, four features etc., to identify the best subsets of features. D’Andrea \emph{et al.} \cite{DANDREA201452} performed a feature selection process using the FFS for elemental analysis on bronze samples. SelectKBest and FFS are easy to implement with available packages such as \textit{sklearn} in python, but when the input variables are high dimensions, SelectKBest and FFS involves a large amount of computation, making it a time-consuming process. \par

Another feature selection technique is Variable Importance (VI), which works similarly to FFS. VI represents the statistical significance of each variable in the data with respect to its effect on the output, which helps weed out certain predictors that contain little information and predominantly extend processing time. VI in Projection (VIP) scores estimate the importance of each variable in the projection and is often used for variable selection. Variables with values significantly less than one are deemed less important and may be good candidates for exclusion from the model. VI and VIP are always used together to select important features. Liang \emph{et al.} \cite{LIANG2020104179} applied VI and VIP to select the feature variables in the LIBS and infrared spectral data to reduce the variable inputs. The VI threshold was optimized based on Out Of Bag (OOB) error estimation. Prasse \emph{et al.} \cite{vrabel2020105872} calculated the VI in the EMSLIBS contest for soil classification. In this study, VI thresholds were applied and the input dimensions were greatly reduced from 40002 to 1379 features. \par

Some researchers have also combined different methods in choosing the best variables. For example, Yan \emph{et al.} \cite{article2} proposed a hybrid variable selection method based on WT and Mean Impact Value (MIV) to extract useful information for calorific value determination of coal. MIV is an indicator to evaluate the importance of independent variables to dependent variables, and the absolute MIV reflects the relative influence degree of the independent variables. WT method is employed to filter the irrelevant information from the broadband LIBS spectra, and then the wavelet coefficients are further eliminated by MIV method. The same authors \cite{YAN201935} proposed another hybrid feature selection method based on V-WSP-PSO which combines the advantages of the variable Wootton, Sergent, PhanTan-Luu's algorithm (V-WSP) \cite{BallabioDavide2014Anvr} based filter method and Particle Swarm Optimization (PSO) based wrapper method. The uncorrelated and redundant features are first eliminated by V-WSP method to form a simplified input subset, and then the retained features are further refined by the PSO method to find a small set of features with high predictive accuracy. \par

Other methods such as handcrafted features, and algorithms such as moving window, Least Absolute Shrinkage and Selection Operator (LASSO) and Genetic Algorithms (GA) have also been tested to select useful information. Handcrafted features like mean, median and standard deviation etc. were used as variable inputs in classifying soils in the EMSLIBS contest \cite{vrabel2020105872}. In this context, another method with a moving window approach was suggested with width and height as variable parameters and the intensity at the peak center (variable at peak maximum) was considered a feature. This corresponded to only 3.2\% of the total data, which greatly reduced the input dimensions and computation time. LASSO is a regression analysis method for both variable selection and regularization to enhance the resulting model’s prediction accuracy and interpretability. In early works, LASSO was applied for feature extraction and multivariate analysis of soil carbon and heavy metals \cite{doi:10.1366/12-06983}\cite{WANG2018300}\cite{osti_1582110}. In GA, a population with a number of individuals is randomly initialized. Followed by fittest-selection, crossover and mutation, the next-generation population is yielded and only those with higher fitness can survive with the generation evolution. Finally, the optimal individual can be filtered from the latest population \cite{LI2021106183}. GA was adopted to select the features to quantify the concentrations of Cu and V in steel with the line-intensity ratios between analyte and matrix (Fe) used as genes \cite{Li:17}. The results indicate that the GA can effectively select features from LIBS to quantify Cu and V in steel. Combining GA and ANN can excellently execute the quantitative analysis for steel samples and further improve analytical accuracy.\par

\subsection{Soil quantitative analysis}
This section gives the details of the quantitative analysis of various soil properties, including macro and trace elements, and other properties such as PH, SOM, contamination, and HF, along with some of the challenges associated with quantification. The quantitative analysis of LIBS is considered its Achilles’ heel because of its complex laser-sample and plasma-particle interactions \cite{Hahn:10}. The earliest applications focused on measurements of major and minor nutrients, such as Carbon (C), Nitrogen (N), Phosphorus (P), and toxic trace elements. Later, LIBS applications are extended to measurements of other soil properties such as PH, SOM, and HF. Unlike other analytical techniques (e.g., ICP-MS), LIBS uses a very small amount of material for analysis (typically $\sim $ ng), hence any changes in the plasma properties will adversely affect the quantification results. In addition, complex chemical compositions of soil samples may lead to serious matrix effects \cite{YangPing2020Dotn}, or if self-absorption is present in the spectral lines, it weakens the peak line intensity \cite{PamuRavi2021CQAo}. Therefore, ML techniques, which can handle the nonlinearity between element concentration and the detection signal, reduce the matrix effect and improve the measurement accuracy, is an excellent candidate for improving elemental analysis in LIBS of soils measurements, which is discussed in detail later in this section.  \par

\subsubsection{Quantification - analyzed soil properties}

 The most common properties evaluated by LIBS are nutrients in soils, including macro- and micro-elements. Guo \emph{et al.} \cite{C9AY00890J} analyzed both macro and micronutrients such as Si, Al, Mg, Ca, Na, K, Mn, Ba, Ti, Cr, Cu, Sr and P in different standard soils to guide the rational use of soil resources. Due to the importance of C, especially organic C (OC), in soil management, agricultural production, and climate change, several reports are available on the analysis of C in soils \cite{https://doi.org/10.2136/sssaj2009.0244}\cite{doi:10.1366/12-06983}\cite{SENESI20167}\cite{Ferreira14}\cite{osti_1582110}. For example, Yongchen \emph{et al.} \cite{2017JApSp..84..731Y} analyzed Mg contents in soils to improve precision fertilization. The fertilizer-related soil pollution is another area covered by LIBS \cite{sun:hal-02276189}. Apart from the nutrients and fertilizer analysis, LIBS has also been used to study the heavy metals in soils. Pb has received special attention among metals due to its high toxicity \cite{article100}. Other heavy element concentrations in agricultural soils such as Cr \cite{Gu2016}, Cu and Zn \cite{Meng:17, C2AY26006A}, Cd \cite{doi:10.1021/acs.analchem.8b01756}, Ca \cite{C8RA07799A, RUHLMANN2018115}, and Ni \cite{WANG2018300} are also monitored. Ferreira \emph{et al.} \cite{FERREIRA201596, Ferreira14} analyzed PH, HD, and SOM in soils using LIBS. Table~\ref{soilQuanIssueTable} depicts the summary of LIBS application in quantitative analysis of soil properties. \par

\begin{table}[htbp]
\caption{LIBS application in soil quantification and analyzed properties with corresponding analytical studies}
\begin{center}
\begin{tabular}{>{\centering\arraybackslash}m{1.7cm}|>{\centering\arraybackslash}m{2cm}|>{\centering\arraybackslash}m{2.7cm}}
\hline
\multicolumn{2}{c|}{\textbf{LIBS application in quantitative }}& \\
\multicolumn{2}{c|}{\textbf{analysis of soil properties}}&\textbf{References} \\

\hline
&Macro Nutrients&\cite{RUHLMANN2018115},\cite{article9},\cite{ilhardt2019119},
\cite{article5},\cite{article8},\cite{lu2018},\cite{C9AY00890J},
\cite{osti_1582110},\cite{Ferreira14},\cite{s20020418},\cite{ELHADDAD201351},
\cite{C7JA00114B},\cite{refId0},\cite{C8RA07799A},\cite{https://doi.org/10.2136/sssaj2009.0244},
\cite{doi:10.1366/12-06983}\\
\cline{2-3}

Soil Nutrients&Trace Elements (micronutrients)&\cite{sun:hal-02276189},\cite{article9},
\cite{NICOLODELLI201970},\cite{s20020418},
\cite{article8},\cite{C9AY00890J},\cite{article5},\cite{VILLASBOAS2016195},
\cite{C7JA00114B},\cite{2017JApSp..84..731Y} \\
\hline
Miscellaneous&PH&\cite{FERREIRA201596},\cite{https://doi.org/10.1111/ejss.12888},\cite{soilsystems3040066}\\
\cline{2-3}
Soil&HF&\cite{Senesi2016LaserbasedSM},\cite{https://doi.org/10.1111/ejss.12888},\cite{Ferreira14}\\
\cline{2-3}
Properties&SOM&\cite{Senesi2016LaserbasedSM},\cite{https://doi.org/10.1111/ejss.12888},\cite{soilsystems3040066},\cite{Ferreira14}\\
\hline
&Fertlizer pollutions&\cite{article200},\cite{sun:hal-02276189}\\
\cline{2-3}
Soil Contamination&Heavy metals&\cite{NAVARRO2008183}, \cite{article100}, \cite{Kwak2012DeterminationOL}, \cite{elhaddad:hal-01025457}, \cite{ZAYTSEV201865}, \cite{doi:10.1177/0003702819826283}, \cite{ELHADDAD201351}, \cite{Gu2016}, \cite{Meng:17}, \cite{C2AY26006A}, \cite{doi:10.1021/acs.analchem.8b01756}, \cite{C8RA07799A}, \cite{RUHLMANN2018115}, \cite{WANG2018300}, \cite{AKHTAR2018143}\\
\hline
Mars Geochemical Surveys&Oxide&\cite{doi:10.1177/0003702816664858},\cite{castorena2021106125},\cite{li2020105850}, \cite{EWUSIANNAN2020105930}\\
\hline
&Soil Geochemical Characteristics&\cite{refId0}\\
\cline{2-3}
Others&Soil Biogeochemical Processes&\cite{TADINI2019454}\\
\hline
\end{tabular}
\label{soilQuanIssueTable}
\end{center}
\end{table}

LIBS has also been used for geochemical surveys of planetary soil and rocks \cite{COUSIN2011805}. The ChemCam remote sensing instrument has investigated many details of the Mars geochemistry and helped researchers make various scientific discoveries \cite{SautterV2014ImaB}. It is also used to analyze geochemical characteristics and biogeochemical processes occurring in the soil in river basins. For example, Chen \emph{et al.} \cite{refId0} conducted element detection and analysis in the Masha River Basin to prevent and control the watershed soil problems. The metals iron (Fe) and aluminum (Al) were analyzed in the Amazon region to study the chemical properties of organic matter to understand the biogeochemical processes occurring in the soil \cite{TADINI2019454}. \par

\subsubsection{ML approaches for quantification}
Various methodological ML approaches including univariant analysis and multivariate analysis are discussed in this section to aid LIBS quantification challenges. Table~\ref{quanMethodTable} provides a summary of the ML methods used in elemental quantification.\par


Univariant analysis (also called the calibration curve method) is a traditional calibration method that establishes calibration curves by relating the detection signal with the concentration of the element to be detected. Certified Reference Material (CRM) and Reference Material (RM) are often preferred to develop calibration curves. However, development of these references is particularly challenging for agricultural applications because of the lack of the specificity of these samples, breadth of reference material required, and variations due to matrix effects \cite{PENG2016260}. The major purpose of univariate analysis is to maximize the SNR and eliminate spectral interference. It is conducted by optimizing the experimental parameters (fluence, lens-to-sample distance, delay, and integration time), which can be achieved by univariate and multivariate experimental designs including orthogonal test design \cite{C3AY41466C}, and chemometrics \cite{NUNES2009565}. However, univariate experimental design usually requires many experimental runs and cannot clarify the relation between different variables, which can be overcome in multivariate experimental methods \cite{PENG2016260}.\par

Unlike univariate analysis, multivariate analysis involves more signal information, which can help deal with matrix effect and shot-to-shot fluctuation and predict element concentrations even in the case of overlapped spectral lines \cite{PENG2016260}. Standard multivariate methods are PCA and PLSR. Over the past decade, the LIBS community has adopted PCA, and the number of scientific articles referring to PCA has steadily increased. Pořízka \emph{et al.} \cite{PORIZKA201865} reviewed the PCA application to LIBS analysis, including data preprocessing, visualization, dimensionality reduction, model building, classification, quantification, and non-conventional multivariate mapping. \par

The standard ML algorithms used for multivariate analysis include SVM and $k$ Nearest Neighbor ($k$-NN). Support Vector Regression (SVR) is a type of SVM for regression issues. The objective of the SVM algorithm is to find a hyperplane in an $N$-dimensional space, where $N$ represents the number of features, that distinctly classifies the data points by maximizing the margin between the closest points in each class. These closest data points are known as the support vectors. Gu \emph{et al.} \cite{Gu2016} built SVR models with whole LIBS spectra and segmented spectra for Cr concentrations in agriculture soils, demonstrating high efficiency and accuracy with SVR on segmented spectra. Guo \emph{et al.} \cite{C9AY00890J} compared the performance of SVR and PLSR on quantitative analysis of soils and this study concluded that the robustness of the SVR model is superior to the PLSR model with a better prediction ability and a lower Relative Standard Deviation (RSD) for both the training data and the test data. The $k$-NN algorithm is a simple, easy-to-implement supervised machine learning algorithm that can solve both classification and regression problems. $k$-NN was used to build a multivariate calibration model for HD and SOM analysis, and the values predicted by the model showed a strong correlation with laser-induced fluorescence spectroscopy (LIFS) results \cite{Ferreira14}. \par

\begin{table}[htbp]
\caption{ML approaches for LIBS quantitative analysis}
\begin{center}
\begin{tabular}{>{\centering\arraybackslash}m{1.3cm}|>{\centering\arraybackslash}m{1.4cm}|>{\centering\arraybackslash}m{1.2cm}|>{\centering\arraybackslash}m{1.8cm}}
\hline
\multicolumn{3}{c|}{\textbf{Method}}&\textbf{References} \\
\hline
\multicolumn{3}{c|}{Univariant}&\cite{Fang2014ElementalAI},\cite{article7},\cite{C2AY26006A},
\cite{doi:10.1080/00387010.2012.747542},\cite{liu:hal-00732080},
\cite{article8},\cite{article9},
\cite{DIAZ:12},\cite{Pareja:13},
\cite{2017JApSp..84..103Y},\cite{doi:10.1021/acs.analchem.6b03969},\cite{doi:10.1177/0003702816664858},
\cite{KIM2013754},\cite{RUHLMANN2018115},\cite{ZAYTSEV201865},
\cite{WANG2018300},\cite{doi:10.1021/acs.analchem.8b01756}
\\
\hline
&&PCA&\cite{PORIZKA201865}, \cite{doi:10.1177/0003702816664858}, \cite{ZAYTSEV201865},\cite{WANG2018300}\\
\cline{3-4}
&Standard Methods&PLSR&\cite{article7},\cite{C9AY00890J},\cite{article5},\cite{soilsystems3040066},
\cite{doi:10.1177/0003702816664858},\cite{osti_1582110},\cite{RUHLMANN2018115},
\cite{Ferreira14},\cite{VILLASBOAS2016195},\cite{s20020418},
\cite{refId0},\cite{WANG2018300},\cite{CLEGG200979},
\cite{doi:10.1366/12-06983},\cite{doi:10.1021/acs.energyfuels.7b01718}
\\
\cline{2-4}
& &MLP&\cite{FERREIRA2011435}\\
\cline{3-4}
& &BPNN&\cite{sun:hal-02276189},\cite{li2020105850},\cite{YAN2017226}\\
\cline{3-4}
&&ANN&\cite{elhaddad:hal-01025457},\cite{ELHADDAD201351}\\
\cline{3-4}
&Neural Network Approaches&KELM&\cite{YAN2017226},\cite{article2},\cite{YAN201935}\\
\cline{3-4}
Multivariate& &CNN&\cite{castorena2021106125},\cite{lu2018},\cite{li2020105850}\\
\cline{2-4}
& &SVR&\cite{C9AY00890J},\cite{Gu2016}, \cite{YAN2017226}\\
\cline{3-4}
&Common ML&LS-SVR&\cite{YAN2017226}\\
\cline{3-4}
&Algorithms&$k$-NN&\cite{Ferreira14}\\
\cline{2-4}
& &NMR&\cite{2017JApSp..84..731Y}\\
\cline{3-4}
& &Internal standard&\cite{Meng:17}, \cite{refId0}\\
\cline{3-4}
& &LASSO&\cite{osti_1582110},\cite{s20020418},\cite{WANG2018300},
\cite{doi:10.1366/12-06983}\\
\cline{3-4}
&Others &MRCE&\cite{osti_1582110},\cite{doi:10.1366/12-06983}\\
\cline{3-4}
& &MLR&\cite{refId0}\\
\cline{3-4}
& &GPR&\cite{s20020418}\\
\hline
\end{tabular}
\label{quanMethodTable}
\end{center}
\end{table}

Neural network approaches such as MLP, ANN, BPNN, Extreme Learning Machine (ELM) and CNN are used in elemental analysis. A neural network is a network or circuit of neurons, and the connections of neurons are modeled as weights. All inputs are modified by weights and summed. These artificial networks may be used for predictive modeling, adaptive control and applications where they can be trained with some dataset. The MLP algorithm is a type of ANN with 3 layers including one hidden layer. If it has more than 1 hidden layer, it is called a deep ANN. CNN is another type of deep learning neural network, which uses kernels to extract the relevant features from the input using the convolution operation. In an early work, Ferreira \emph{et al.} \cite{FERREIRA2011435} compared MLP to a linear fit method for the quantitative analysis. The linear fit method chooses the emission lines to correlate their intensities with the reference concentration in the linear fit calibration procedure. The values determined by MLP showed lower prediction errors, accuracy and lower limits of detection. ANN was applied for on-site quantitative analysis by Haddad \emph{et al.}\cite{ELHADDAD201351} and concluded that ANN's prediction ability is improved with the combination of target analyte emission lines and the extra lines from other chemical elements inside the studied samples. ANN was further investigated by the same research group \cite{elhaddad:hal-01025457} for on-site quantitative analysis of lead in soils. Sun \emph{et al.} \cite{sun:hal-02276189} developed a multivariate model based on BPNN and concluded that the resulting multivariate model provided great improvements compared to the univariate one.\par

Among the neural network approaches, ELM proposed by Huang \emph{et al.} \cite{HUANG2006489} has a fast learning speed and good generalization performance. When the feature mapping function of hidden neurons is unknown, the kernel function can be used to improve the algorithm's stability \cite{6035797}, which is called Kernel-based Extreme Learning Machine (KELM). The KELM  model \cite{YAN2017226}\cite{article2}\cite{YAN201935} demonstrated excellent performance compared with SVR, Least Squares-SVR (LS-SVR) and BPNN in carbon and sulfur content determination in coal, which indicates that LIBS combined with KELM method is a promising technique for real-time online, rapid analysis in the coal industry.\par

The time-resolved LIBS has also been analyzed by CNNs, which extracts and integrates the information of both the wavelength and time dimension, showing the improvement in the detection of K in soil \cite{lu2018}. Li \emph{et al.} \cite{li2020105850} compared the performance of CNN with two alternative schemes based on BPNN and PLSR. A spectral CNN operation on LIBS signals is proposed to (1) disentangle spectral signals from the sources of sensor uncertainty (i.e., pre-process) and (2) get qualitative and quantitative measures of the chemical content of a sample \cite{castorena2021106125}.\par

Several authors have utilized different multivariate regression methods for elemental quantitative analysis. For example, Bricklemyer \emph{et al.} \cite{doi:10.1366/12-06983} used two regression shrinkage and variable selection approaches, LASSO and sparse Multivariate Regression with Covariance Estimation (MRCE), for soil C analysis and the identification of wavelengths that are important for C prediction. MRCE constructs a sparse estimator of a multivariate regression coefficient matrix that accounts for the correlation of the response variables, which involves penalized likelihood with simultaneous estimation of the regression coefficients and the covariance structure \cite{doi:10.1198/jcgs.2010.09188}. This work demonstrated that LASSO and MRCE approaches improve calibration prediction accuracy over PLSR, but requires additional testing with increased soil and target analyte diversity. Bricklemyer's group also reported soil C analysis with three multivariate regression approaches (LASSO, MRCE and PLS2) \cite{osti_1582110} and noticed MRCE provided the best C prediction accuracy. Simultaneously, Wang \emph{et al.} \cite{WANG2018300} proposed multivariate methods, using LASSO and Principal Component Regression (PCR), for heavy metal analysis and found they effectively reduced the matrix interference compared with univariant analysis. In a recent study, soil nutrient was detected with PLSR, LASSO and Gaussian Process Regression (GPR) \cite{s20020418}. GPR is a nonparametric, Bayesian approach that works well on small datasets and can provide uncertainty measurements on the predictions. \par

Internal standard and Nonlinear Multivariate Regression (NMR) methods have also been applied for multivariate analysis with promising performance. Methods such as Multi Linear Regression (MLR), PLSR, and internal standards have been used to improve the fit of the calibration curve and reduce the quantitative analysis error for element analysis of watershed in soils \cite{refId0}. The internal standard method was also used to detect heavy metal elements and improve LIBS data stability by 6\% \cite{Meng:17}. The NMR method was applied to analyze Mg contents in soil with LIBS \cite{2017JApSp..84..731Y}.

\subsection{Soil Classification}

This section reviews the development of LIBS in soil classification, including the issues solved by soil discrimination and the fusion of LIBS and ML approaches in classifying soils.

\subsubsection{Analyzed properties with soil classification}

In agriculture, discriminating soil types and contaminants is very crucial for crop productivity and LIBS is widely used  for this application. Kim \emph{et al.} \cite{KIM2013754} applied LIBS to discriminate between contaminated soils with heavy metals or oils and clean soils. Similarly, Haddad \emph{et al.} \cite{elhaddad:hal-01025457} applied LIBS to detect Pb contamination. In addition to soil discrimination based on different contaminants, LIBS has also been applied to classify soil types and textures \cite{article4, kaiser2020}. Classifying soil textures help quantify its nutrients. For example, Rühlmann \emph{et al.} \cite{RUHLMANN2018115} classified soils roughly into sand, silt, and clay to exclude the outliers first. Then the filtered LIBS spectra was used to quantify the nutrients. Conversely, the determination of soil nutrients can also help with soil texture classification. For instance, Villas-Boas \emph{et al.} \cite{VILLASBOAS2016195} classified and estimated the proportions of soil textures (sand, clay, and silt) based on the assumption of the chemical compositions, including the content of Si, Na, Fe, Ti, Ca, K, Al, Co, Mg, V, Ba, and Be.  \par

Apart from soil discrimination in agriculture, there has been an increased interest in the forensic applications of soil examination. Provenance determination can be accomplished by elemental analysis of major, minor, and trace components to create a signature/profile for each soil type and location, which can be compared to a database to assist with locating and identifying the scene of crimes \cite{article9, fc924bbeae0c4571b3c33fc9af5e1a33}. Table~\ref{soilClaasificationIssueTable} provides a literature survey of LIBS for soil classifications. \par

\begin{table}[htbp]
\caption{LIBS for soil classification}
\begin{center}
\begin{tabular}{>{\centering\arraybackslash}m{1.8cm}|>{\centering\arraybackslash}m{2cm}|>{\centering\arraybackslash}m{1.4cm}}
\hline
\multicolumn{2}{c|}{\textbf{LIBS Classification based on soil properties}}&\textbf{References} \\
\hline
&Lead contamination&\cite{doi:10.1177/0003702819826283},\cite{elhaddad:hal-01025457}\\
\cline{2-3}
Soil Contamination &Heavy metals contamination&\cite{Meng:17}\\
\cline{2-3}
&Heavy metals and oil contamination&\cite{KIM2013754}\\
\hline
\multicolumn{2}{c|}{Soil types}&\cite{article4},\cite{vrabel2020105872}\\
\hline
\multicolumn{2}{c|}{Soil textures (sand, clay, silt)}&\cite{RUHLMANN2018115},\cite{VILLASBOAS2016195}\\
\hline
\multicolumn{2}{c|}{Soils from different sites}&\cite{article9},\cite{fc924bbeae0c4571b3c33fc9af5e1a33}\\
\hline
\end{tabular}
\label{soilClaasificationIssueTable}
\end{center}
\end{table}

\subsubsection{ML techniques used in soil classification}

\begin{figure}[htbp]
\centerline{\includegraphics[width=\columnwidth]{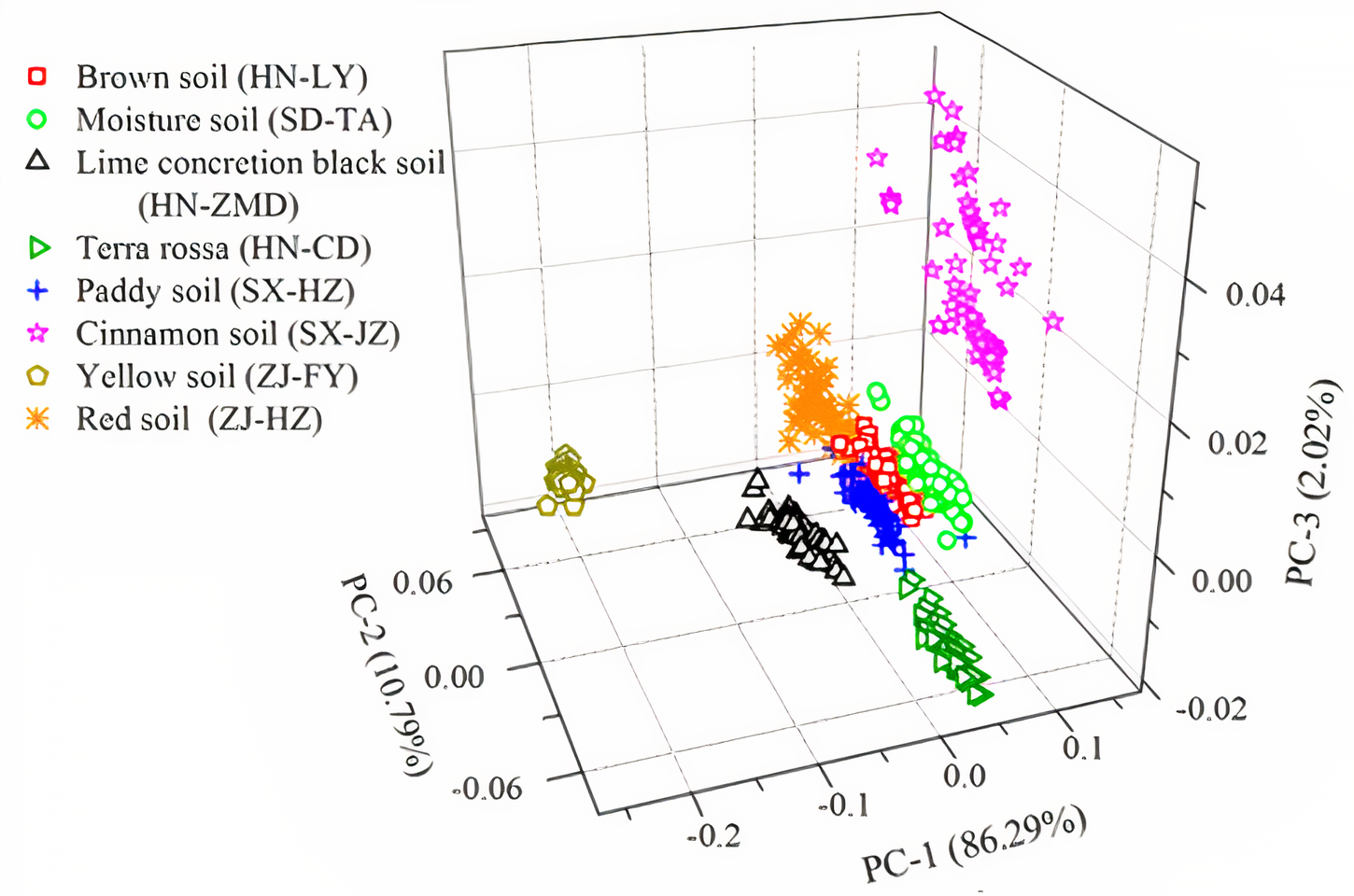}}
\caption{The score plot of first three PCs from PCA on LIBS data of eight types of soil samples \cite{article4}.}
\label{soilclassification_fig}
\end{figure}

This section attempts to summarize the methods used in soil classification with a fusion of LIBS and ML techniques. PCA is a popular method to discriminate soils \cite{article9}\cite{Meng:17}\cite{RUHLMANN2018115}\cite{vrabel2020105872}. When the output is categorical, PLS can be used to discriminate the input variables, which is called PLS Discriminant Analysis (PLS-DA) and is a commonly used method in soil classification (\cite{KIM2013754}\cite{doi:10.1177/0003702819826283}\cite{vrabel2020105872}). LDA is similar to PCA and has been applied to distinguish soils \cite{doi:10.1021/acs.energyfuels.6b02279}\cite{GAZELI2020125329}\cite{vrabel2020105872}. An example of soil classification using PCA is shown in Figure~\ref{soilclassification_fig}, where eight types of soil samples (Brown, Moisture, Black, Terra Rosa, Paddy, Cinnamon, Yellow, and Red Soil) are classified based on LIBS spectra. The various ML tools used for classification are summarized in Table~\ref{classMethodTable}.\par

\begin{table}[htbp]
\caption{Survey of ML tools for LIBS Classification}
\begin{center}
\begin{tabular}{>{\centering\arraybackslash}m{1.5cm}|>{\centering\arraybackslash}m{2cm}|>{\centering\arraybackslash}m{1.5cm}}
\hline
\multicolumn{2}{c|}{\textbf{Method}}&\textbf{References} \\
\hline
&PCA&\cite{RUHLMANN2018115},\cite{Meng:17},\cite{vrabel2020105872},\cite{article9}, \cite{HUANG2022106451}\\
\cline{2-3}
Standard Methods&PLS-DA&\cite{doi:10.1177/0003702819826283},\cite{vrabel2020105872},\cite{KIM2013754}\\
\cline{2-3}
&LDA&\cite{doi:10.1021/acs.energyfuels.6b02279},\cite{vrabel2020105872},\cite{GAZELI2020125329}\\
\hline
&SVM&\cite{doi:10.1177/0003702819826283},\cite{CHEN2020105801},\cite{https://doi.org/10.1002/cem.2422}\\
\cline{2-3}
Common ML&LS-SVM&\cite{article4}\\
\cline{2-3}
Algorithms&RF&\cite{LIANG2020104179}\\
\cline{2-3}
&$k$-NN&\cite{CHEN2020105801}\\
\hline
&SOM algorithm&\cite{PAGNOTTA201570},\cite{Pagnotta2017TheDA},\cite{TANG2018179}\\
\cline{2-3}
Neural Network&ANN&\cite{elhaddad:hal-01025457},\cite{vrabel2020105872},\cite{CHEN2020105801}\\
\cline{2-3}
Approaches&ICA-WNN&\cite{C7JA00218A}\\
\cline{2-3}
&CNN&\cite{CHEN2020105801}\\
\hline
&SIMCA&\cite{article4},\cite{2018Classification}\\
\cline{2-3}
Others&LIBS spectra peak ratios&\cite{fc924bbeae0c4571b3c33fc9af5e1a33}\\
\cline{2-3}
&Moving window algorithm&\cite{doi:10.1021/acs.analchem.7b04124}\\
\hline
\end{tabular}
\label{classMethodTable}
\end{center}
\end{table}

Random Forest (RF) and SVM are the most common ML techniques employed in classification with LIBS. SVM was applied to classify suspect powders \cite{https://doi.org/10.1002/cem.2422}, which performs well for the data and setting presented and could be used in other cases where one of the classes has some known elemental structure. Least Squares SVM (LS-SVM) can be used to find the solution by solving a set of linear equations instead of a convex Quadratic Programming (QP) problem for classical SVMs and is a class of kernel-based learning methods \cite{article4}. The classification performance of LS-SVM was compared with Soft Independent Modelling by Class Analogy (SIMCA), which demonstrated that the LS-SVM model is the optimal model for discriminating different types of soils \cite{article4}. SIMCA is a statistical method for the supervised classification of data, in which the term soft refers to the fact that the classifier can identify samples as belonging to multiple classes and not necessarily produce a classification of samples into non-overlapping classes. SIMCA was applied to classify minerals \cite{2018Classification}, which suggests that it is efficient in identifying a wide variety of geological samples with similar sample types.  \par 

Because of its easy use without hyper-parameter tuning, RF is one of the most used algorithms in classification and regression. RF is an ensemble learning method by constructing a multitude of decision trees with a training dataset. Liang \emph{et al.} \cite{LIANG2020104179} applied RF on fusion data of LIBS and infrared spectra for the classification and discrimination of compound salvia miltiorrhiza. RF was also used in an EMSLIBS contest in 2019 for feature extraction and soil classification \cite{vrabel2020105872}. \par

Neural network approaches are introduced in recent works such as SOM algorithm, ANN, and CNN \cite{TANG2018179}\cite{doi:10.1177/0003702819826283}\cite{CHEN2020105801}. SOM algorithm has been applied in material identification by several groups \cite{PAGNOTTA201570, TANG2018179}. SOM algorithm is a type of ANN and is trained with unsupervised learning to reduce dimensions. Pagnotta \emph{et al.} \cite{PAGNOTTA201570} used the SOM algorithm to identify different metallic alloys with similar compositions and later extended their study to realize qualitative identification of the mineralogical composition in Roman mortars based on micro-LIBS detection \cite{Pagnotta2017TheDA}. Tang \emph{et al.} \cite{TANG2018179} applied SOM algorithm combined with $k$-means to classify 20 industrial polymers. Their results indicated that unsupervised learning algorithms is an effective method for classifying industrial polymers, which has practical applications in environmental conservation and resource recycling. \par

Haddad \emph{et al.} \cite{elhaddad:hal-01025457} presented the efficacy of ANN in predicting the type of soil matrix by calculating three values related to the silicate, calcareous, and ores poles. Zhang \emph{et al.} \cite{C7JA00218A} compared the performance of the Independent Component Analysis-Wavelet Neural Network (ICA-WNN) with ANN in coal ash classification, which demonstrated that the ICA-WNN model has better classification performance than the ANN model. Deep learning approaches combined with common ML techniques are demonstrated to be efficient in soil discrimination. For example, Zhao \emph{et al.} \cite{doi:10.1177/0003702819826283} combined deep learning with the Deep Belief Network (DBN) and SVM to classify soil contamination and concluded that the proposed fusion method yields better performance than PLS. \par

CNN was applied for lithological recognition, and the classification performance of 2D CNN is compared to 1D CNN and other optimized machine learning models, including: $k$-NN, PCA-$k$-NN, SVM, PCA-SVM, PLS-DA and HA-ANN (Human Assisted ANN) \cite{CHEN2020105801}. This study demonstrated that the 2D CNN costs less training time than the 1D CNN and achieves the best prediction accuracy compared to other methods, which presents great potential for lithological recognition. Woods \emph{et al.} \cite{fc924bbeae0c4571b3c33fc9af5e1a33} used peak ratios of soil elements to discriminate the specimens using a three-sigma criterion (average value ${\pm}$ three times the standard deviation), resulting in 92.4\% discrimination accuracy. Moros \emph{et al.} \cite{doi:10.1021/acs.analchem.7b04124} employed a moving window algorithm to categorize rocks from the Mars surface. The great majority of forecasts have matched the real identities of the inspected targets. 

\subsection{Self-absorption}

Self-absorption/self-reversal effects should be considered when analyzing LIBS spectra because they can distort the spectral line profile by increasing the line width and reducing the line intensity. This eventually affects the quantification and classification analysis \cite{fu2019detecting}. In self-absorption, the photon emitted by the atom is reabsorbed by the same kind of atom and this process is prominent for resonance or near-resonance transitions. Self-absorption distorts the line profile because the absorption is strongest at the line center and weakest at the wings of the spectral profile. However, evaluating the amount of self-absorption in a LIBS plume is challenging by simply observing the lineshape \cite{YiRongxing2016Iots}. \par



The self-absorption of emission lines causes nonlinearity in the univariate regression model and is most often seen for strong lines terminating at the ground or near the ground state when the analyte concentration is high \cite{TAKAHASHI201731}. Hou \emph{et al.} \cite{ HOU_2019}  summarized methods used to correct the deformed spectra caused by self-absorption, including Curve of Growth (COG), spectra fitting and modeling, and self-absorption coefficient. Different ML algorithms are introduced in multivariant analysis, such as ANN for overcoming the nonlinearity caused by self-absorption. Rezaei \emph{et al.} \cite{Rezaei_article} applied self-absorption correction for high concentrations species and found that the ANN predictions are more accurate than the calibration curve results. Takahashi \emph{et al.} \cite{ TAKAHASHI201731} studied the effects of self-absorption with PCR, PLS and ANN, in which they mentioned that the effects of self-absorption can be alleviated with PCR and PLS by reducing their weights, and ANNs can reduce these effects by modeling the non-linear relationship using a flexible statistical model. The feature extraction process can also reduce the influence of self-absorption by selecting features located in the wings of a line profile instead of line center, which reduces the influence of self-absorption significantly \cite{Sun_2021}. \par

\section{Prospects}
\label{prospects}

LIBS has already achieved remarkable results in the analysis of soils because of its ability to provide nearly non-destructive, simultaneous multi-elemental analysis in real-time. The combination of LIBS with ML techniques has many prospects in improving its the technique for many applications. Key prospects are listed below: 

\begin{itemize}
\item \textbf{Portable LIBS:} With the availability of compact and portable lasers and spectrometers, the miniaturized field-portable LIBS instrumentation with precision and accuracy is feasible for next-generation soil analysis of large-scale fields at a relatively low cost.   
\item \textbf{Whole soil application:} The emerging application of the LIBS technique to evaluate the HD of SOM in whole soil samples is very promising. However, further research efforts are required to improve the sensitivity, selectivity, accuracy, and precision \cite{Senesi2016LaserbasedSM}. 
\item \textbf{Improvement in LOD:} A number of novel experimental have been used for improving the sensitivity, such as fs LIBS \cite{C5JA00301F, 2020-LIBSbook-Hari}, Nanoparticle-Enhanced (NE) LIBS \cite{DEGIACOMO201419} \cite{Ohta:09}, Dual-pulse LIBS \cite{article5}\cite{CORSI2006748}\cite{NICOLODELLI201523}\cite{Nicolodelli:18}, LIBS assisted by LIF (LIBS-LIF) \cite{Gao:18}\cite{Nicolodelli:18}\cite{2021-JAP-LIz-LIF} and Microwave-Assisted (MA)-LIBS \cite{liu:hal-00732080}. Combining LIBS emission enhancement methods and ML techniques for improving  the LOD will be useful.  
\item \textbf{Automatic feature engineering and selection:} Deep learning techniques are able to conduct feature selection automatically in the training stage, which speeds up the data analysis and makes LIBS more attractive for soil analysis. 
\item \textbf{Efficient dimensionality reduction:} ML techniques are able to handle high dimensional data efficiently with dimensionality reduction algorithms, which remove redundant features and noises and accelerate the data processing. 
\item \textbf{Image-assisted matrix correction:} Real-time imaging of the plasma plume is helpful for correcting the matrix effect \cite{image-assisted-matrix}. With the fast development of computer vision and ML in image processing, ML techniques combined with image-assisted LIBS becomes a promising technology and will greatly promote the development of LIBS in soil analysis.  
\item \textbf{Fusion of ML and LIBS imaging:} Image analysis assisted by ML techniques makes LIBS imaging promising in understanding the complex chemical nature of root-soil interactions \cite{ilhardt2019119}.
\end{itemize}

\section{Summary}
\label{conclusions}
LIBS is a technique with great potential for soil chemical and physical characterization due to its simple setup, cost-effectiveness and rapid (few-second) analysis time per sample. This review provides an overview of the LIBS with an emphasis on soil analysis, challenges, and the prospects of ML tools to advance the technique. Various models and algorithms combined with ML techniques are reviewed for soil classification and elemental analysis. The combination of ML and LIBS can provide improved quantitative capability for the advanced realization of LIBS for soil analysis in the future.

\section*{Acknowledgment}

We acknowledge financial support from Alberta Innovates through the Smart Food and Agriculture Digitization Challenge (Agreement No. 202100740, Hussein) and the Natural Sciences and Engineering Research Council of Canada (Grant No. RGPIN-2021-04373, Hussein). Pacific Northwest National Laboratory is a multi-program national laboratory operated by Battelle for the U.S. Department of Energy under Contract DE-AC05-76RL01830.

\newacronym{ann}{ANN}{Artificial Neural Networks}
\newacronym{libs}{LIBS}{Laser-Induced Breakdown Spectroscopy}
\newacronym{ml}{ML}{Machine Learning}
\newacronym{icp-ms}{ICP-MS}{Inductively Coupled Plasma-Mass Spectrometry}
\newacronym{faas}{FAAS}{Flame Atomic Absorption Spectrometry}
\newacronym{icp-oes}{ICP-OES}{Inductively Coupled Plasma-Optical Emission Spectroscopy}
\newacronym{hd}{HD}{Humidification Degree}
\newacronym{som1}{SOM}{Soil Organic Matter}
\newacronym{lod}{LOD}{Limit of Detection}
\newacronym{bpnn}{BPNN}{Back-Propagation Neural Network}
\newacronym{rbfnn}{RBFNN}{Radial Basis Function Neural Networks}
\newacronym{som2}{SOM}{Self-Organizing Map}
\newacronym{cnn}{CNN}{Convolutional Neural Networks}
\newacronym{ai}{AI}{Artificial  Intelligence}
\newacronym{haz}{HAZ}{Heat-Affected Zone}
\newacronym{ccd}{CCD}{Charge-Coupled Device}
\newacronym{iccd}{ICCD}{Intensified Charge-coupled Device}
\newacronym{nist}{NIST}{National Institute of Standards and Technology}
\newacronym{lif}{LIF}{Laser-Induced Fluorescence}
\newacronym{mpls}{MPLS}{Morphological Weighted Penalized Least Squares}
\newacronym{snv}{SNV}{Standard Normal Variate}
\newacronym{msc}{MSC}{Multiplicative Scatter Correction}
\newacronym{mc}{MC}{Mean Centring}
\newacronym{pc}{PC}{Principal Component}
\newacronym{pca}{PCA}{Principal Component Analysis}
\newacronym{pls}{PLS}{Partial Least Squares}
\newacronym{lda}{LDA}{Linear Discriminant Analysis}
\newacronym{emslibs}{EMSLIBS}{Euro-Mediterranean Symposium on Laser-Induced Breakdown Spectroscopy}
\newacronym{ipls}{iPLS}{interval PLS}
\newacronym{mipw-pls}{mIPW-PLS}{modified Iterative Predictor Weighting-PLS}
\newacronym{mlp}{MLP}{Multilayer Perception}
\newacronym{anova}{ANOVA}{Analysis of Variance}
\newacronym{ffs}{FFS}{Forward Feature Selection}
\newacronym{vi}{VI}{Variable Importance}
\newacronym{oob}{OOB}{Out Of Bag}
\newacronym{vip}{VIP}{Variable Importance in Projection}
\newacronym{miv}{MIV}{Mean Impact Value}
\newacronym{pso}{PSO}{Particle Swarm Optimization}
\newacronym{lasso}{LASSO}{Least Absolute Shrinkage and Selection Operator}
\newacronym{ga}{GA}{Genetic Algorithms}
\newacronym{cf}{CF}{Calibration-Free}
\newacronym{cf-libs}{CF-LIBS}{Calibration-Free Laser-Induced Breakdown Spectroscopy}
\newacronym{bsa}{BSA}{Binary Search Algorithm}
\newacronym{crm}{CRM}{Certified Reference Material}
\newacronym{rm}{RM}{Reference Material}
\newacronym{plsr}{PLSR}{Partial Least Squares Regression}
\newacronym{svr}{SVR}{Support Vector Regression}
\newacronym{svm}{SVM}{Support Vector Machine}
\newacronym{knn}{$k$-NN}{$k$ Nearest Neighbor}
\newacronym{rsd}{RSD}{Relative Standard Deviation}
\newacronym{elm}{ELM}{Extreme Learning Machine}
\newacronym{kelm}{KELM}{Kernel-based Extreme Learning Machine}
\newacronym{ls-svr}{LS-SVR}{Least Squares-Support Vector Regression}
\newacronym{mrce}{MRCE}{Multivariate Regression with Covariance Estimation}
\newacronym{gpr}{GPR}{Gaussian Process Regression}
\newacronym{nmr}{NMR}{Nonlinear Multivariate Regression}
\newacronym{mlr}{MLR}{Multi Linear Regression}
\newacronym{ls-svm}{LS-SVM}{Least Squares Support Vector Machine}
\newacronym{qp}{QP}{Quadratic Programming}
\newacronym{simca}{SIMCA}{Soft Independent Modelling by Class Analogy}
\newacronym{rf}{RF}{Random Forest}
\newacronym{pls-da}{PLS-DA}{Partial Least Squares Discriminant Analysis}
\newacronym{ica-wnn}{ICA-WNN}{Independent Component Analysis-Wavelet Neural Network}
\newacronym{dbn}{DBN}{Deep Belief Network}
\newacronym{pcr}{PCR}{Principal Component Regression}
\newacronym{gpr}{GPR}{Gaussian Process Regression}
\newacronym{nmr}{NMR}{Nonlinear Multivariate Regression}
\newacronym{libs-lif}{LIBS-LIF}{Laser-Induced Breakdown Spectroscopy assisted by Laser-Induced Fluorescence}
\newacronym{ma-libs}{MA-LIBS}{Microwave-Assisted Laser-Induced Breakdown Spectroscopy}
\newacronym{v-wsp-pso}{V-WSP-PSO}{variable - Wootton, Sergent, PhanTan-Luu's algorithm - Particle Swarm Optimization}
\newacronym{v-wsp}{V-WSP}{variable - Wootton, Sergent, PhanTan-Luu's algorithm}
\newacronym{pso}{PSO}{Particle Swarm Optimization}
\newacronym{snr}{SNR}{Signal-to-Noise Ratio}
\newacronym{wt}{WT}{Wavelet Transformation}
\newacronym{sbr}{SBR}{Signal to Background Ratio}
\newacronym{sfr}{SFR}{Full Spectrum Normalization}
\newacronym{scr}{SCR}{Carbon Normalization}
\newacronym{oc}{OC}{Organic Carbon}
\newacronym{knn}{$k$-NN}{$k$-Nearest Neighbour}
\newacronym{lifs}{LIFS}{Laser-Induced Fluorescence Spectroscopy}
\newacronym{ha-ann}{HA-ANN}{Human Assisted - Artificial Neural Networks}
\newacronym{cog}{COG}{Curve of Growth}
\newacronym{fs}{fs}{femtosecond}
 
\glsaddall
\printglossary[type=\acronymtype,title=Acronyms]

\bibliographystyle{IEEEtran}

\bibliography{LIBS_survey}
\end{document}